\documentclass[iop]{emulateapj}

\usepackage[T1]{fontenc}
\usepackage{natbib}
\usepackage{bm}
\usepackage[breaklinks,colorlinks,citecolor=blue]{hyperref}
\usepackage[all]{hypcap}
\usepackage{bm}
\usepackage{color}
\usepackage[dvipsnames]{xcolor}
\usepackage{setspace}
\usepackage{graphicx}
\usepackage{amsmath}

\renewcommand{\arraystretch}{0.6}

\shorttitle{}
\shortauthors{}

\begin{document}

\title{The Impact of Radio AGN Bubble Composition on the Dynamics and Thermal Balance of the Intracluster Medium}
%\title{(Dissecting) Heating Mechanisms of AGN Feedback in Cool-Core Galaxy Clusters}

\author{H.-Y.\ Karen Yang\altaffilmark{1,2},
Massimo Gaspari\altaffilmark{3,*},
Carl Marlow\altaffilmark{4}} 
\altaffiltext{1}{Department of Astronomy, University of Maryland, College Park, MD, USA} 
\altaffiltext{2}{Joint Space-Science Institute, College Park, MD, USA} 
\altaffiltext{3}{Department of Astrophysical Sciences, Princeton University, Princeton, NJ, USA}
\altaffiltext{4}{Department of Physics and Astronomy, University of Alabama, Huntsville, AL, USA}
\altaffiltext{*}{{\it Einstein} and {\it Spitzer} Fellow}
\email{Email: hsyang@astro.umd.edu}

\begin{abstract}

Feeding and feedback of active galactic nuclei (AGN) are critical for understanding the dynamics and thermodynamics of the intracluster medium (ICM) within the cores of galaxy clusters. While radio bubbles inflated by AGN jets could be dynamically supported by cosmic rays (CRs), the impact of CR-dominated jets are not well understood. In this work, we perform three-dimensional simulations of CR-jet feedback in an isolated cluster atmosphere; we find that CR jets impact the multiphase gas differently than jets dominated by kinetic energy. In particular, CR bubbles can more efficiently uplift the cluster gas and cause an outward expansion of the hot ICM. Due to adiabatic cooling from the expansion and less efficient heating from CR bubbles by direct mixing, the ICM is more prone to local thermal instabilities, which will later enhance chaotic cold accretion onto the AGN. The  amount of cold gas formed during the bubble formation and its late-time evolution sensitively depend on whether CR transport processes are included or not. We also find that low-level, subsonic driving of turbulence by AGN jets holds for both kinetic and CR jets; nevertheless, the kinematics is consistent with the {\it Hitomi} measurements. Finally, we carefully discuss the key observable signatures of each bubble model, focusing on gamma-ray emission (and related comparison with {\it Fermi}), as well as thermal Sunyaev-Zel'dovich constraints.
\end{abstract}


%===================================================

\section{Introduction}\label{intro}

% Importance of AGN feeding and feedback in clusters

Energetic outputs from the supermassive black holes (SMBHs) are influential for the evolution of galaxies and galaxy clusters. In clusters with short central radiative cooling times, or cool-core (CC) clusters, it is generally believed that feedback from the central active galactic nucleus (AGN) is what maintains the global thermal balance within the cluster cores and prevents the clusters from catastrophic cooling, motivated by the prevalence of radio bubbles or X-ray cavities inflated by AGN jets in CC clusters, and the correlation between cavity power and cooling luminosity for observed bubbles \citep[][for a review]{McNamara12}. However, the detailed processes of feeding and feedback are still debated due to the large separation of scales and complex physics involved. 

% Recent advances in the field

In recent years, there has been a substantial advancement in the understanding of AGN feedback in clusters (and massive galaxies), thanks to high-resolution numerical simulations including increasingly complex physics. While earlier simulations based on Bondi accretion of hot gas and accretion-powered energy injections showed some limited self-regulation of SMBH growth \citep{Sijacki07, Cattaneo07, Dubois10, Yang12a}, more realistic hydrodynamic simulations including chaotic cold-gas accretion (CCA -- \citealt{Gaspari13, Gaspari15, Gaspari17}; see also \citealt{Pizzolato05}) and momentum-driven jets/outflows have better reproduced the positive temperature gradient and thermodynamical properties of the condensed multiphase gas observed in CC systems \citep{Gaspari11, Gaspari12, Gaspari13, Li14, Prasad15, Yang16a, Yang16b, Prasad17, Martizzi18}. In terms of SMBH feeding, CCA has gained increasing support through a variety of multiwavelength observations \citep{Werner14, Voit15, Tremblay16, Tremblay18, David17, Lakhchaura18, Maccagni18, McDonald18, Gaspari18, Temi18}. The detailed processes of cold-gas formation due to local thermal instabilities have also benefited from local idealized simulations \citep{McCourt12, Sharma10, Sharma12, Meece15, Meece17}, as well as analytical works \citep{McNamara16,Voit17}. 

Regarding feedback, the heating mechanisms proposed include cavity heating \citep{Churazov01, Bruggen03}, weak shocks \citep{Fabian03, Nulsen05, Gaspari11, Randall15, Li17}, sound waves \citep{Ruszkowski04a, Ruszkowski04b, Fabian05, Zweibel17b, Fabian17}, thermal conduction \citep{Zakamska03, Voigt04, Yang16a}, turbulent dissipation \citep{Dennis05, Zhuravleva14}, turbulent/direct mixing between ultra-hot bubbles and the ICM \citep{David01, Kim03, Gaspari15x, Hillel16, Yang16b}, and cosmic rays (CRs) \citep{Guo08, Mathews08, Pfrommer13, R17, Weinberger17, Bourne17, Jacob17a, Jacob17b, Ehlert18}. Recent hydrodynamic simulations have also provided a more comprehensive picture and allows one to determine the relative importance among the different mechanisms. Specifically, for processes that can be probed by purely hydrodynamic simulations, direct mixing \citep{Hillel16, Yang16b} and shock heating \citep{Barai14, Barai16,Li17} are more likely to be the primary heating mechanisms, whereas turbulent dissipation seems to be subdominant \citep{Reynolds15, Yang16b, Hitomi16, Fabian17, Bambic18}. Regardless of the heating mechanism, it is noted by \cite{Yang16b} that heating is not required to balance cooling exactly throughout the cluster cores, nor do the jets need to heat isotropically. Instead, fluid motions in a process of `gentle circulation' would self-adjust to transport and compensate heat provided by the AGN jets.  

% But bubble composition is uncertain, and studies including CRs are rather sparse

Despite the substantial progress, one of the biggest issues is that the composition of the radio bubbles is largely unknown. Observations of cluster radio bubbles suggest that the primary pressure support of many radio bubbles does not come from radio-emitting CR electrons, but has to be magnetic field, ultra-hot thermal plasma, or non-radiating CR protons \citep{Dunn04, DeYoung06, Birzan08, Croston18}. While thermal bubbles, which are naturally produced by kinetic-energy-dominated jets because of efficient thermalization by shocks during bubble inflation, have been studied extensively as mentioned above, studies on magnetic-field-dominated jets \citep[e.g.,][]{Li06, ONeill10, Sutter12} and CR-dominated jets in the cluster context are relatively sparse in the literature. 

% Review literature of CR feedback

If cluster radio bubbles are dominated by CR protons, the feeding and feedback of the AGN could be widely different from thermal bubbles. First, the ICM containing the CR fluid is less dense and more buoyant due to the extra pressure support from CRs. Indeed, early simulations of CR feedback found that CR bubbles, instead of providing heat to the intracluster medium (ICM), have a net cooling effect by efficient driving of a mass outflow. Second, CR jets (which are typically internally subsonic) could more easily generate `fat' bubbles as seen near the center of the Perseus cluster \citep{Guo11, Guo15, Guo16}. Furthermore, energy transfer from the CRs to the gas via Alfv\'en waves through the streaming instability \citep{Kulsrud69, Wentzel74, Zweibel13} is a viable heating mechanism in CC clusters \citep{Guo08, Pfrommer13, R17, Jacob17a, Jacob17b, Ehlert18}. While CR transport mechanisms (e.g., diffusion and streaming) are demonstrated to be crucial \citep[e.g.,][]{R17}, they are governed by microphysical plasma processes which are not well understood until more recently \citep[e.g.,][]{Wiener13, Zweibel17, Wiener18}. These all point to the necessity of detailed investigations of CR feedback. 

% What we do: 3D to study turbulence, high-resolution sims of single outburst to bring insights to heating/cooling, observable signatures to distinguish kinetic vs. CR bubbles.

To this end, we perform three-dimensional (3D) hydrodynamic simulations of CR-jet feedback in a Perseus-like cluster. We investigate the detailed evolution of a single AGN outburst in order to study the impact of CR-dominated jets on the dynamics and thermal balance of the ICM. We contrast CR jets with kinetic jets and compare models with and without CR transport processes. Our simulations are carried out in 3D, a key improvement compared with the previous two-dimensional simulations \citep{Mathews08,Guo11}, allowing  us to quantify the properties of ICM turbulence and to accurately probe line-of-sight observables. Our study of a single AGN outburst greatly reduces the complexity of multiple injections and allows us to gain insights into the physical processes at play, an approach complementary to global simulations of self-regulated CR feedback \citep{R17}. Finally, we seek to make contact with observations and find distinct observable signatures for thermal vs.\ CR bubbles that may be used to inform future observational studies. 

The structure of the paper is as follows. In \S\ref{sec:method}, we summarize the equations and assumptions about CR physics in \S\ref{sec:cr} and describe the simulation setups in \S\ref{sec:sim}. In \S\ref{sec:results}, we present results on the general evolution (\S\ref{sec:evol}), heating and cooling processes (\S\ref{sec:hc}), generation of turbulence (\S\ref{sec:turb}), and observable signatures in gamma rays (\S\ref{sec:gamma}) and thermal Sunyaev-Zel'dovich (SZ) effect (\S\ref{sec:sz}). The last section is included in light of the recent tentative detection of AGN bubbles via the thermal SZ effect \citep{Abdulla18}. We summarize our findings in \S\ref{sec:conclusion}.

%===================================================
\section{Methodology}
\label{sec:method}

We carry out 3D hydrodynamic simulations of a single AGN jet injection in an idealized Perseus-like cluster using the adaptive-mesh-refinement (AMR) code FLASH \citep{Flash, Dubey08}. In this paper we focus on the comparisons of three cases: kinetic energy-dominated jets, CR-dominated jets without CR transport, and CR-dominated jets with CR transport. The setups for the kinetic-energy- and CR-dominated jets (hereafter referred to as kinetic jets and CR jets, respectively) are similar to those in \cite{Yang16b} and \cite{R17}, respectively, though in the current work we examine the detailed evolution of one jet-inflated bubble rather than the long-term evolution of the cluster (e.g., \citealt{GS17}). Motivated by the observational constraints from \cite{Dunn04} that many radio bubbles could be energetically dominated by non-radiating relativistic particles, we assume the CRs in our simulations to be purely CR protons. As we will show in \S~\ref{sec:gamma}, this assumption is consistent with constraints obtained from the gamma-ray upper limits of clusters. Also, for the current work we neglect the effects of magnetic field in order to avoid introducing a large parameter space (due to magnetic field orientations, coherence lengths, etc) and focus on the differences with and without CRs. Since the magnetic field could affect integrity of the bubbles and CR propagation, we will investigate its effects in future work. We refer the readers to the above references for details. In the following we give a brief summary and highlight the differences compared to our previous works.

\subsection{Cosmic-ray physics}
\label{sec:cr}

In the simulations we treat CRs as a second fluid and solve the following hydrodynamic equations including the effects of CRs \citep{Yang12b, R17}:
\begin{eqnarray}
&& \frac{\partial \rho}{\partial t} + \nabla \cdot (\rho {\bm v}) = 0,\\
&& \frac{\partial \rho {\bm v}}{\partial t} + \nabla \cdot \left( \rho {\bm v}{\bm v} \right) + \nabla p_{\rm tot} = \rho {\bm g},\\
&& \frac{\partial e}{\partial t} + \nabla \cdot \left[ (e+p_{\rm tot}){\bm v} \right] = \rho {\bm v} \cdot {\bm g} + \nabla \cdot ({\bm \kappa} \cdot \nabla e_{\rm cr}) + {\mathcal H_{\rm cr}},\quad \\
&& \frac{\partial e_{\rm cr}}{\partial t} + \nabla \cdot (e_{\rm cr} {\bm v}) = -p_{\rm cr} \nabla \cdot {\bm v} + \nabla \cdot ({\bm \kappa} \cdot \nabla e_{\rm cr}) + {\mathcal C_{\rm cr}},
\label{eq:hydro}
\end{eqnarray}
where $\rho$ is the gas density, ${\bm v}$ is the gas velocity, ${\bm g}$ is the gravitational field, ${\bm \kappa}$ is the CR diffusion tensor, $e_{\rm cr}$ is the CR energy density, and $e=(1/2)\rho v^2 + e_{\rm th} + e_{\rm cr}$ is the total energy density (including kinetic, thermal, and CR energy). The total pressure is $p_{\rm tot} = (\gamma -1)e_{\rm th} + (\gamma_{\rm cr} -1) e_{\rm cr}$, where $e_{\rm th}$ is the internal energy density of the gas, $\gamma=5/3$ is the adiabatic index for ideal gas, and $\gamma_{\rm cr}=4/3$ is the effective adiabatic index of CRs. ${\mathcal H_{\rm cr}}$ represents the rate of change of total energy density due to hadronic CR losses, and ${\mathcal C_{\rm cr}}$ is the CR cooling rate due to Coulomb, hadronic, and streaming processes. 

The above equations include CR advection, CR diffusion, dynamical effects from CR pressure, and CR heating to the thermal gas. The underlying assumption of this formalism is the `extrinsic turbulence' model of CR propagation \citep{Zweibel17}, in which CRs are scattered by waves that are part of a turbulent cascade in the background plasma. Under the condition of balanced turbulence, the CRs essentially advect with the gas because the transport due to forward-propagating and backward-propagating waves is canceled out, and there is no collisionless heating of the gas due to energy transfer via Alfv\'en waves associated with the streaming instability \citep{Kulsrud69, Wentzel74}. These considerations justify the simple treatment of CR transport as a combination of advection and diffusion, and therefore this formalism is widely adopted by early simulations including CRs \citep[e.g.,][]{Guo08, Mathews08}. 

In this extrinsic turbulence picture, CR diffusion with respect to the {\it mean} field can exist due to field-line wandering as well as diverging field lines due to shearing by the Alfv\'en modes in the turbulence. For high-$M_A$ turbulence ($M_A\equiv v_L/v_A$, where $M_A$ is the Alfv\'enic Mach number, $v_L$ is the turbulent velocity at the injection scale $L$, and $v_A$ is the Alfv\'en speed), which is largely valid in the ICM \citep{Miniati15}, one could estimate the diffusion coefficient to be $\kappa_\perp = \kappa_\parallel \sim (1/3)\,l_A v$ 
%\MG{[MG: Again, I don't understand the 1/3 factor, it seems ad hoc, please specify the physical meaning in the text. KY: This is directly quoted from Eq.\ 24 of Yan \& Lazarian (2008). Would you like to keep this factor or remove it since this is an order-of-magnitude estimate?]} MAX: let's keep it for now
\citep{Yan08}, where $l_A=L/M_A^3$ is the scale at which the turbulent velocity is equal to the Alfv\'en speed, and $v$ is the gas velocity since in the extrinsic picture the CRs are advected with the thermal fluid. For typical values in the ICM, $L \sim 1$ Mpc, $v_L\sim 1000$ km s$^{-1}$, $M_A \sim 10$, $l_A \sim 1$ kpc, $v \sim 100$ km s$^{-1}$, and $\kappa \sim 10^{28}$ cm$^2$ s$^{-1}$, which is close to the canonical value often adopted for the Galaxy. For most of our simulations, we assume the extrinsic turbulence model and use a constant value of $\kappa = 3 \times 10^{28}$ cm$^2$ s$^{-1}$.    

There can be collisional heating of gas due to Coulomb and hadronic interactions. The CR energy loss rates due to the Coulomb and hadronic processes can be written as \citep{YoastHull13, R17}
\begin{equation}
{\mathcal C_{\rm cr,c}} = -4.93 \times 10^{-19} \frac{n-4}{n-3} \frac{e_{\rm cr}\rho}{E_{\rm min}} \frac{\rho}{\mu_{\rm e}m_{\rm p}} {\rm erg\ cm^{-3}\ s^{-1}}
\end{equation}
and
\begin{equation}
{\mathcal C_{\rm cr,h}} = -8.56 \times 10^{-19} \frac{n-4}{n-3} \frac{e_{\rm cr}\rho}{E_{\rm min}} \frac{\rho}{\mu_{\rm p}m_{\rm p}} {\rm erg\ cm^{-3}\ s^{-1}},
\end{equation}
where $n>4$ is the slope of the CR distribution function in momentum, $E_{\rm min}$ is the minimum energy of CRs, and $\mu_{\rm e}$ and $\mu_{\rm p}$ are the mean molecular weights per electron and proton, respectively. Note that all the CR energy loss due to Coulomb collisions is transferred to the gas, but only a small fraction ($\sim 1/6$, see \citealt{Mannheim94}) of the inelastic energy goes into secondary electrons and can be used to heat the gas \citep{Guo08} and the remainder is removed as gamma-ray emission and neutrinos in the pion production process. Therefore, the rate of change of the total energy density, which includes the thermal and CR energy densities, is ${\mathcal H_{\rm cr}} = (5/6) {\mathcal C_{\rm cr,h}} < 0$, and the CR energy density loss rate is ${\mathcal C_{\rm cr}} = {\mathcal C_{\rm cr,c}} + {\mathcal C_{\rm cr,h}}$.

\begin{table*}[tp]
\caption{Summary of simulations}
\begin{center}
\begin{tabular}{ccccc}
\hline
\hline
Run & $f_{\rm cr}$ & CR diffusion & Collisional heating & Collisionless heating \\ 
\hline
KIN & 0.001 & no & no & no \\
CR & 0.9 & no & no & no \\
CRh & 0.9 & no & yes & no \\
CRdh & 0.9 & yes & yes & no \\
CRdhs & 0.9 & yes & yes & yes \\
\hline
\end{tabular}
\end{center}
\label{tbl:sims}
\end{table*}

Alternatively, one could consider the `self-confinement' picture of CR transport, in which the CR transport speed is limited by self-excited Alfv\'en waves via the streaming instability \citep{Kulsrud69, Wentzel74, Zweibel13}. Since simulating the effect of streaming is numerically challenging \citep[][but see recent computational advancements by \citealt{Jiang18} and \citealt{Thomas18}]{Sharma10}, some of the previous works have utilized approximations to simplify the equations \citep[e.g.,][]{Sharma09, Ehlert18} instead of directly simulating CR streaming \citep{R17}. For instance, under the assumption that the CRs are strongly scattered (so that CRs effectively travel with $v_A$ along the field lines) and that there are tangled magnetic structures on small scales, one could show that the CR transport can be approximated as spatial diffusion \citep{Sharma09, Ehlert18}, with $\kappa \sim l_B v_A \sim 3\times 10^{28} (l_B/{\rm kpc})(v_A/100\ {\rm km\ s^{-1}})$ cm$^2$ s$^{-1}$, where $l_B$ is the characteristic scale of magnetic field tanglement. Since this treatment is formally identical to the above equations except that there would be collisionless CR heating due to the streaming instability in addition to the collisional Coulomb and hadronic heating, we include one simulation that explores the effect of CR streaming. Specifically, for this simulation we use a constant diffusion coefficient of $\kappa = 3 \times 10^{28}$ cm$^2$ s$^{-1}$ (which is consistent with a parallel diffusion coefficient $\kappa_\parallel = 10^{29}$ cm$^2$ s$^{-1}$ as used in the magnetohydrodynamic simulation of \cite{Ehlert18} assuming efficiently tangled field), but there is additional CR energy loss due to streaming, ${\mathcal C_{\rm cr,s}}=-{\bm v}_A \cdot \nabla P_{\rm cr}$. Note that because this energy is transferred from the CRs to the gas, the rate of change of the total energy due to streaming is zero (i.e., ${\mathcal H_{\rm cr,s}} = 0$). Since the simulation is purely hydrodynamic, when computing the streaming heating term, we assume a constant magnetic field of 1 $\mu$G.

The simulations performed in this study is summarized in Table \ref{tbl:sims}. The three simulations that will be compared in detail are KIN, CR, and CRdh. But to investigate the effects of CR transport, we also include one simulation with no CR transport but only Coulomb and hadronic heating (CRh), and one simulation with diffusion as well as heating due to Coulomb, hadronic, and streaming (CRdhs).

\subsection{Simulation setup}
\label{sec:sim}

The simulation cube is 500 kpc on a side and is adaptively refined on steep temperature gradients up to an AMR refinement level of 8, which corresponds to a resolution element of 0.5 kpc.\footnote{We ran additional simulations with a peak resolution of 0.25kpc and found that mixing and the associated heating are more efficient in the higher-resolution simulations. However, our main conclusions, namely, the qualitative differences between KIN, CR, and CRdh simulations, remain unaffected.} The reflecting boundary condition is used,
%\MG{[MG: what happens if we use outflowing boundaries? Reflecting in the middle of the cluster is a bit unrealistic, are we sure we do not get waves back from the boundaries? KY: If outflow boundaries were used, the total mass and energy would not be strictly conserved, which would substantially complicates the interpretation of the energy evolution as discussed in Figure 3.]} MAX: OK agreed
which is chosen in order to conserve the total energy within the simulation domain for the purpose of studying the evolution of different energy components (\S\ref{sec:hc}). We verified that the waves resulted from materials bounced back from the reflecting boundaries cause negligible velocity perturbations ($\sim 1$ km s$^{-1}$) and do not interfere with the forward shock induced by the AGN jets within the duration of the simulations (100 Myr). For the default simulations presented in this work, radiative cooling is turned off so that the total energy is conserved after the initial energy injection by the jets. We also ran an additional set of simulations including radiative cooling in order to more accurately quantify the amount of cold gas formed from local thermal instabilities. For the radiative simulations, cooling is computed using the tabulated table of \cite{SutherlandDopita} assuming 1/3 solar metallicity. The gas profiles of the cluster are initialized using empirical fits to the observed Perseus cluster assuming hydrostatic equilibrium within a static Navarro-Frenk-White (NFW) \citep{Navarro96} gravitational potential. 

The simulated AGN outburst has a duration of 10 Myr and total jet power of $\dot{E_{\rm ej}}=5\times 10^{45}$ erg s$^{-1}$, which is approximately the average jet power obtained in our previous simulation of self-regulated feedback
%\MG{[MG: I suggest to add also that this is a typical AGN power observed in massive clusters, e.g., from X-ray cavities (e.g., Birzan et al. 2012)]}
for the same setup of the Perseus cluster \citep{Yang16b}. A fraction of the total jet energy is injected as CRs ($f_{\rm cr}$) and the remainder is injected as kinetic energy ($1-f_{\rm cr}$). The CR fraction $f_{\rm cr}$ is chosen to be 0.001 and 0.9 for the kinetic and CR jets, respectively. 
The kinetic jets have a small but nonzero CR fraction so that the jet fluid is dyed. The injected mass and momentum by the AGN jets can be expressed as $\dot{M}_{\rm ej}=2(1-f_{\rm cr}) \dot{E}_{\rm ej}/v_{\rm ej}^2$ and $\dot{P}_{\rm ej}=\dot{M}_{\rm ej} v_{\rm ej}$, respectively. To make a fair comparison, we require the kinetic and CR jets to have the same momentum, which gives the constraint that $\dot{M}_{\rm ej} v_{\rm ej} = 2(1-f_{\rm cr}) \dot{E}_{\rm ej} / v_{\rm ej}$ is a constant, or $v_{\rm ej} \propto 1-f_{\rm cr}$. To satisfy this requirement, we choose $v_{\rm ej}=0.0999c$ and 0.01c for kinetic and CR jets, respectively ($c$ is the speed of light). These jet speeds represent jets that have already gone through rapid deceleration on kpc scales \citep{Laing06}. We also note that the parameters chosen here are comparable to previous works \citep[e.g.,][]{Guo08}.
The feedback is applied to a cylinder with radius of 2 kpc and height of 4 kpc, and the bipolar jets are injected along the $\pm z$ axis in the simulation domain.  

%===================================================

\section{Results}
\label{sec:results}
 
\subsection{Bubble evolution} 
\label{sec:evol}

\begin{figure*}[tbp]
\begin{center}
\includegraphics[scale=0.7]{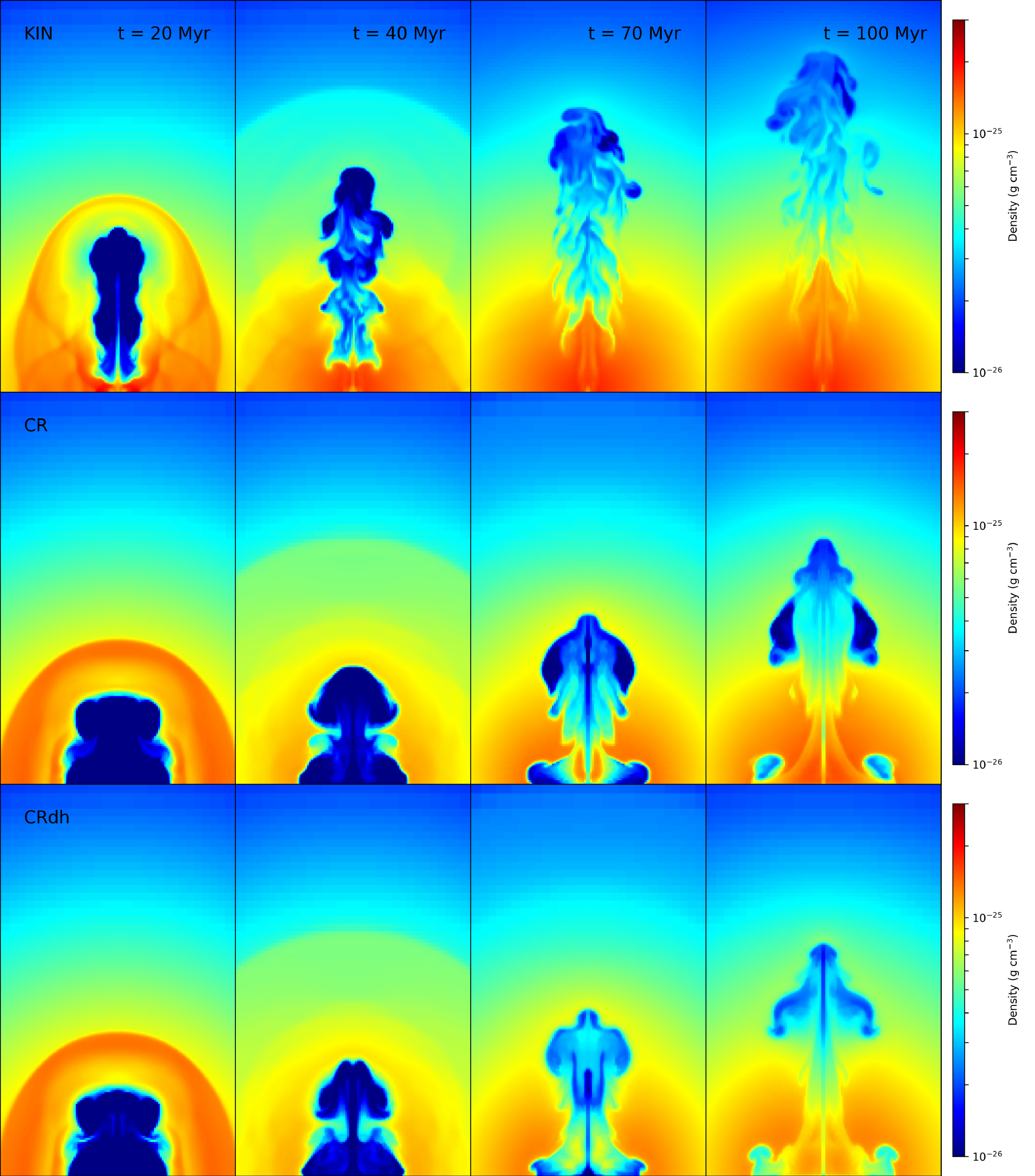} 
\caption{Evolution of the gas density for simulations of kinetic-energy-dominated jets (KIN; top row), CR-dominated jets with no diffusion or heating (CR; middle row), and CR-dominated jets with diffusion and collisional heating (CRdh; bottom row). The physical scale for each panel is 60 kpc by 100 kpc.} %\MG{[MG: the CRdhs map is missing here and in the next 2 figures, please add it. KY: Figure 1-3 for the CRdhs run is qualitatively very similar to the CRdh run, which is expected since both include CR transport and heating. For this reason, I think plotting these three cases (KIN, CR, and CRdh) here would better illustrate the differences between kinetic jets versus CR jets, and between cases with and without transport. The CRdh and CRdhs runs differ only in the quantitative sense, e.g., the amount of cold gas formed in the simulations, which is why I include the result in Figure 4.]}} MAX: OK
\label{fig:dens}
\end{center}
\end{figure*}

Figure \ref{fig:dens} shows the evolution of gas density for the three representative simulations, i.e., KIN, CR, and CRdh. At the early stage of the evolution, the bubbles form from the lateral expansion when the ram pressure of the jets balances the external pressure. Weak shocks (with Mach number of $\sim 1.2$ and $\sim 1.05$ at $t=10$ Myr for kinetic and CR jets, respectively) are driven as a result of the initial supersonic expansion during bubble formation. As the jets are turned off at $t=10$ Myr, the bubbles are detached from the base of the jets and rise outward due to buoyancy forces. Due to the density contrast and velocity shear at the interface between the bubbles and the shocked ambient gas, the bubbles are subject to Rayleigh-Taylor (RT) and Kelvin-Helmholtz (KH) instabilities and gradually mixed with the surrounding ICM, reducing the density contrast toward the end of the simulations. While bubbles inflated by both kinetic and CR jets are shredded eventually, the bubbles in the KIN run are disrupted on a somewhat shorter KH timescale due to higher jet velocity and larger velocity shear at the bubble-ICM interface (see below).

\begin{figure*}[tbp]
\begin{center}
\includegraphics[scale=0.6]{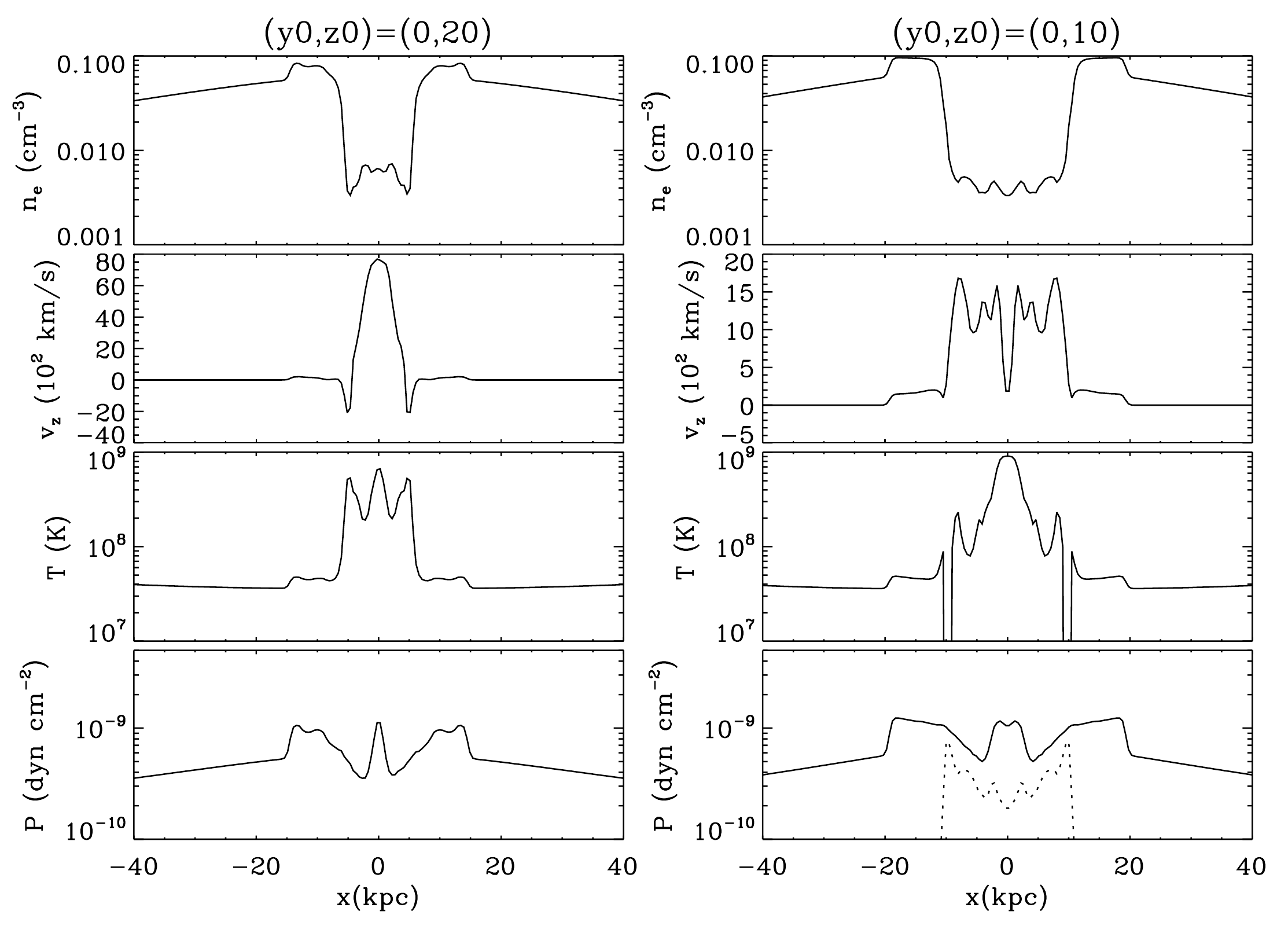} 
\caption{Horizontal profiles for the KIN (left) and CR (right) simulations at $t=12$ Myr. Panels from top to bottom show profiles of electron number density, vertical velocity, gas temperature, and pressure (solid lines represent the total pressure from both gas and CRs, and the dotted line is the CR pressure). The profiles are evaluated at the intercepts $y=y_0$ and $z=z_0$, where $y_0=0$ and the value of $z_0$ roughly correspond to the height of the bubbles at $t=12$ Myr.}
\label{fig:prfs}
\end{center}
\end{figure*}

To take a closer look, Figure \ref{fig:prfs} shows horizontal profiles of electron number density, vertical velocity, gas temperature, and pressure (for the total pressure and CR pressure) for the KIN (left column) and CR (right column) cases. The profiles are evaluated at $t=12$ Myr, which is right after the jets are turned off, and at intercepts $y=y_0$ and $z=z_0$, where $y_0=0$ and the chosen $z_0$ roughly corresponds to the height of the bubbles at $t=12$ Myr. The bubbles (the region with low density) in both cases have electron number density of $\sim (4-7)\times 10^{-3}$ cm$^{-3}$ and $T \sim 10^8-10^9$ K. There are dips in the gas temperature in the CR case close to the bubble surface, indicating formation of cold clumps (we will discuss them in more detail later). The pressure within the bubbles is $\approx 4\times 10^{-10}$ to $10^{-9}$ dyne cm$^{-2}$ for both cases. However, while in the KIN case the bubbles are supported almost purely by thermal pressure, in the CR case the bubble pressure is contributed roughly equally by thermal gas and CRs. As we will discuss later, this difference in the amount of thermal energy contained with bubbles is important for the thermodynamic evolution of the ICM (see \S~\ref{sec:hc}).

We could also calculate for the KIN and CR cases the timescales for the growth of the RT and KH instabilities respectively:
\begin{eqnarray}
t_{\rm RT} &\sim& 18.2\ {\rm Myr} \sqrt{\frac{1+\eta}{1-\eta}} \left( \frac{g}{3\times 10^{-8}} \right)^{-1/2} \left( \frac{\lambda}{20\ {\rm kpc}} \right)^{1/2},\\
t_{\rm KH} &\sim& 10\ {\rm Myr} \left( \frac{\lambda}{20\ {\rm kpc}} \right) \left( \frac{\Delta v}{10^3\ {\rm km\ s}^{-1}} \right)^{-1} \left( \frac{\eta}{0.1} \right)^{-1/2},
\end{eqnarray}
where $\eta$ is the density contrast across the bubble-ICM interface, $g$ is the magnitude of gravitational acceleration in {\it cgs} units, $\lambda$ is the wavelength of the mode of instability under consideration, and $\Delta v$ is the shear velocity at the bubble-ICM interface. For both cases, the timescale for the growth of the RT instability is $\sim 20.1$ Myr for a density contrast of $\eta \sim 0.1$. Indeed, this is roughly the time when distortions start to appear at the top of the bubbles (see Figure \ref{fig:dens}). The KH instability sets in even earlier, which manifests itself as ripples on the side of the bubbles. At $t=12$ Myr, the shear velocity across the bubble surface is $\sim 3000$ km s$^{-1}$ for the KIN simulation and $\sim 1500$ km s$^{-1}$ in the CR simulation. These translate into $t_{\rm KH} \sim$ 6.7 and 3.3 Myr for the KIN and CR cases, respectively. Note that although the shear velocity at the bubble surface is not much greater in the KIN case than in the CR case at $t=12$ Myr, at later times their difference becomes greater as the CR bubbles are decelerated more significantly (see discussion below). This is why the KH features are more prominent in the KIN case in the late-time evolution of the bubbles.

% Fig: For a certain time, snapshots of density, temperature, CR energy density, projected X-ray emissivity

\begin{figure*}[tbp]
\begin{center}
\includegraphics[scale=0.8]{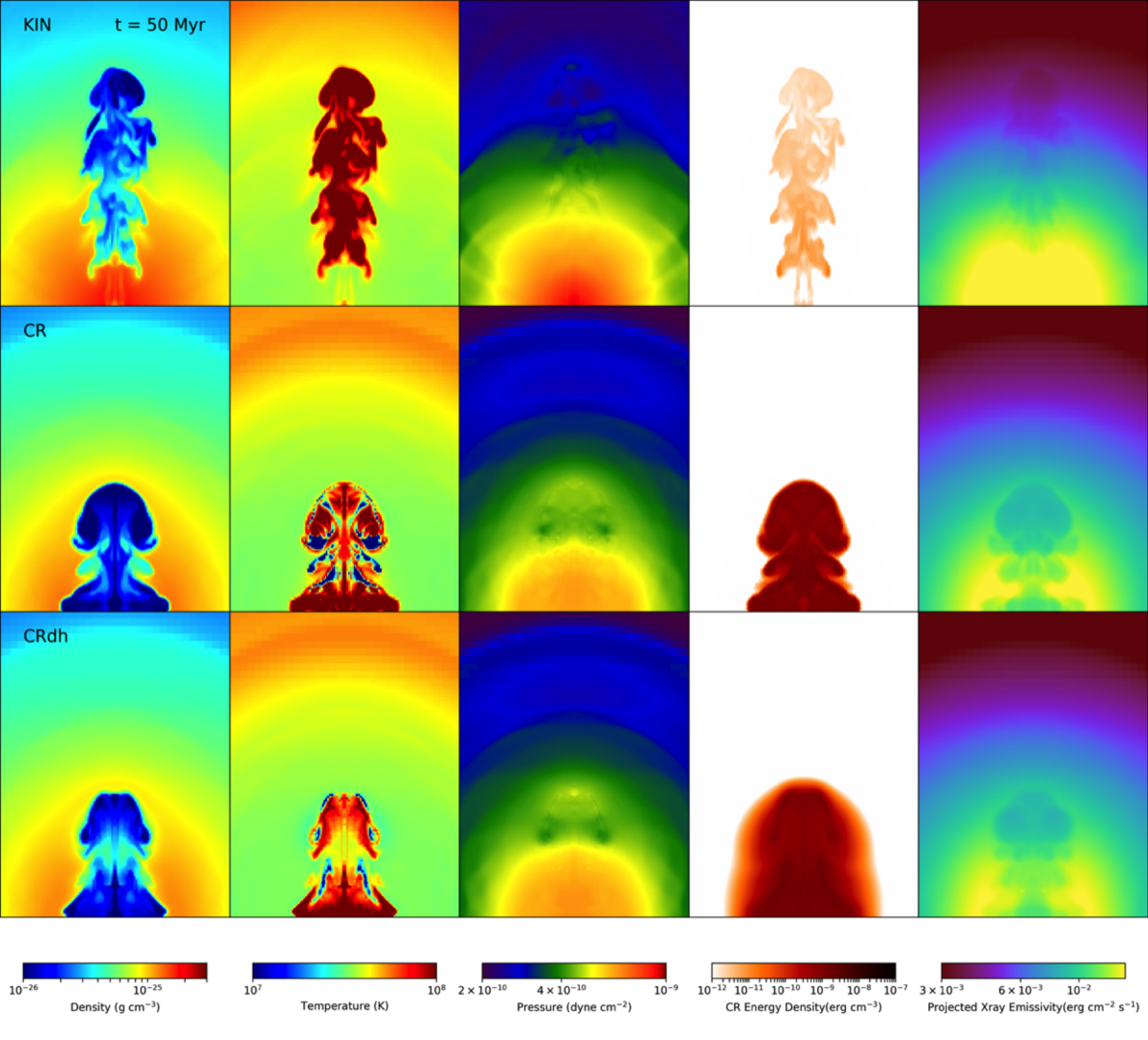} 
\caption{From left to right are slices of gas density, temperature, total pressure (including thermal and CR pressures), CR energy density, and projected X-ray emissivity at $t=50$ Myr. Top to bottom rows show results for the KIN, CR, and CRdh simulations, respectively. The physical scale for each panel is 60 kpc by 80 kpc.}
\label{fig:slices}
\end{center}
\end{figure*}

Figure \ref{fig:slices} shows the gas density, temperature, total pressure, CR energy density, and projected X-ray emissivity at $t=50$ Myr for the three simulations. For simulations without CR transport (KIN and CR), the CRs are well confined within the cavities, whereas for the run with transport (CRdh), the distribution of CRs extends to a much larger region beyond the bubble edges. From the maps of projected X-ray emissivity, one can see that one of the most noticeable differences between the kinetic-jet- and CR-jet-inflated bubbles is their morphology, namely, the former is more elongated and the latter is `fatter', consistent with previous findings \citep{Guo11, Guo15}. Because the CR jets are slower, the initial pressure contrast between the jets and the ambient medium causes the lateral expansion of the bubbles more prominent than the kinetic jets. Subsequently, the initially fatter bubbles in the CR case are more significantly decelerated due to larger surface areas, which makes the bubbles even fatter in the later expansion. Note that this morphological difference would appear whenever one confronts internally supersonic jets versus internally subsonic jets (where the internal Mach number is defined to be $v_{\rm ej}$/$c_{\rm s,jet}$ and $c_{\rm s, jet}$ is the sound speed within the jets at the time of injection) \citep{Guo15}. 
%\MG{[MG: Indeed; let's show the maps of the Mach number too! KY: Note that the internal Mach number is defined for the jets `at injection', i.e., as initial parameters, and therefore would not be the same as the Mach number plotted after the jets propagate on the simulation grid and get decelerated. Therefore I think plotting the Mach number would not help illustrate this point. But if instead you'd like to know the Mach number of the shock generated, I now quote the number in the beginning of the section.]} MAX: OK good
Our results are consistent with this conclusion because, for realistic jet parameters, kinetic jets are more likely internally supersonic and CR jets are internally subsonic because the effective sound speed of the composite gas plus CR fluid is typically very high. 

Another important difference between kinetic and CR jets is that, while the bubbles contain mainly hot thermal gas from the thermalization of jet kinetic energy, cold gas (defined as gas with $T<5\times 10^5$ K in our simulations) is formed in the two CR simulations due to local thermal instabilities (see second column in Figure \ref{fig:slices}). Simply from visual comparison, one could see that the amount of cold gas is different for the simulations with and without CR transport. The formation of cold gas was not discussed in previous simulations of CR feedback in clusters \citep[e.g.,][]{Guo11, Ehlert18}. However, because it is crucial for understanding the heating and cooling processes in cluster cores and the CCA feeding of the central SMBH (see \S\ref{intro}), we will discuss them in more detail in the following section.

\subsection{Heating and cooling}
\label{sec:hc}

In this section, we investigate the impact of bubble composition on the thermodynamics of the ICM, in particular, the heating and cooling processes with the cluster core. We show the evolution of the hot and cold phases of the ICM and how the kinetic and CR jets affect them very differently. 

% Fig: Energetics vs. t

\begin{figure*}[tbp]
\begin{center}
\includegraphics[scale=0.65]{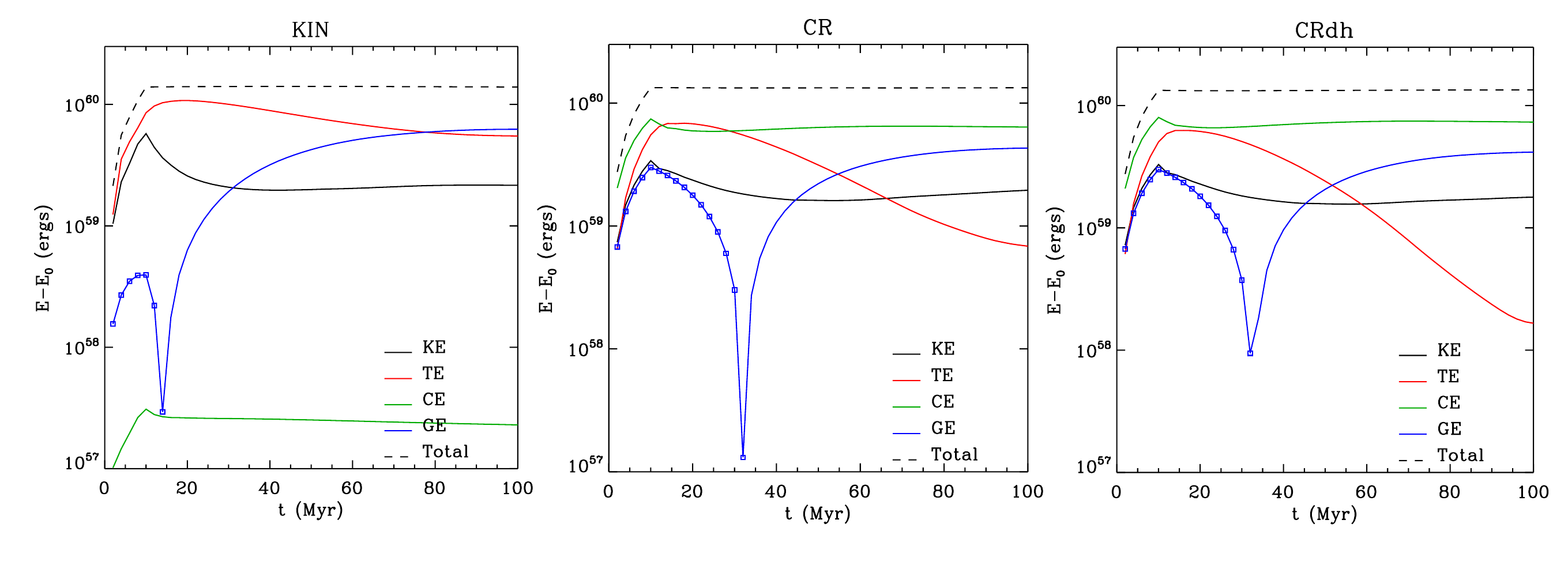} 
\caption{Evolution of different energy components (KE: kinetic energy, TE: thermal energy, CE: CR energy, GE: gravitational potential energy) with respect to their initial values for the KIN (left), CR (middle), and CRdh (right) cases. The dashed lines represent the sums of all components. Negative values are plotted using open symbols.}
\label{fig:energies}
\end{center}
\end{figure*}

Figure \ref{fig:energies} shows the evolution of different energy components within the simulation domain, including kinetic, thermal, CR, and gravitational energies, for the KIN, CR, CRdh cases. All energies are subtracted by their initial values to focus on their temporal evolution. The summation of all four energy components (dash lines) reflects an initial rise due to the energy injection by the AGN jets before $t=10$ Myr and is conserved afterwards. Negative values are plotted using open squares. 

For the KIN simulation, because of the initial thermalization of the jet kinetic energy by shocks, at the end of AGN injection at $t=10$ Myr, most of the injected kinetic energy becomes thermal energy stored within the bubbles as well as the weak shocks. 
The thermal energy then gradually decreases with time owing to adiabatic expansion of the cluster atmosphere (note that eventually the gas will fall back due to gravity; however, this has not occurred during the simulated period of time). The CR energy in the KIN simulation remains negligible in terms of dynamics. The change in gravitational energy is defined as $E_{\rm g}(t)-E_{\rm g}(t=0)=\sum_i \rho_i(t) \phi_i dV_i - \sum_i \rho_{i}(t=0) \phi_i dV_i$, where the summation is performed over all cells with index $i$, $\phi<0$ is the NFW potential, and $dV$ is the volume of a grid cell. It decreases during the active phase of AGN injection because of the mass injection with the jets, i.e., $\rho(t)>\rho(t=0)$ whereas $\phi$ and $dV$ are constant in time. The absolute value is non-negligible because the injected mass is located close to the cluster center where the potential well is the deepest. After $t=10$ Myr, $E_g(t)-E_g(t=0)$ steadily increases with time and eventually becomes positive due to the expansion of the cluster atmosphere. The kinetic energy after the initial thermalization is shared by the bubbles and the weak shocks. It is always subdominant throughout the simulation. In \S\ref{sec:turb} we will discuss the ICM kinematics in more detail. 

The evolution of energies are similar for the CR and CRdh simulations (middle and right panels in Figure \ref{fig:energies}, respectively). Similar to the KIN case, the kinetic energy remains subdominant over the course of the evolution. The initially injected mass close to the cluster center causes the change in gravitational energy to be negative (and more so because the CR-dominated bubbles rise more slowly compared to the KIN case) and increases after the jets are turned off as the cluster gas expands. The CR energy remains the dominant component throughout the simulations. During the initial stage of bubble formation, part of the injected CR energy is lost due to adiabatic expansion. Afterwards, the CRs do little work to the surroundings and hence the total CR energy is roughly conserved. The lost CR energy becomes thermal energy shared between the weak shocks and the bubbles. Compared to the KIN case, the decrease in thermal energy due to adiabatic losses after $t=10$ Myr is much more dramatic, indicating more significant expansion of the cluster gas.   

\begin{figure}[tbp]
\begin{center}
\includegraphics[scale=0.9]{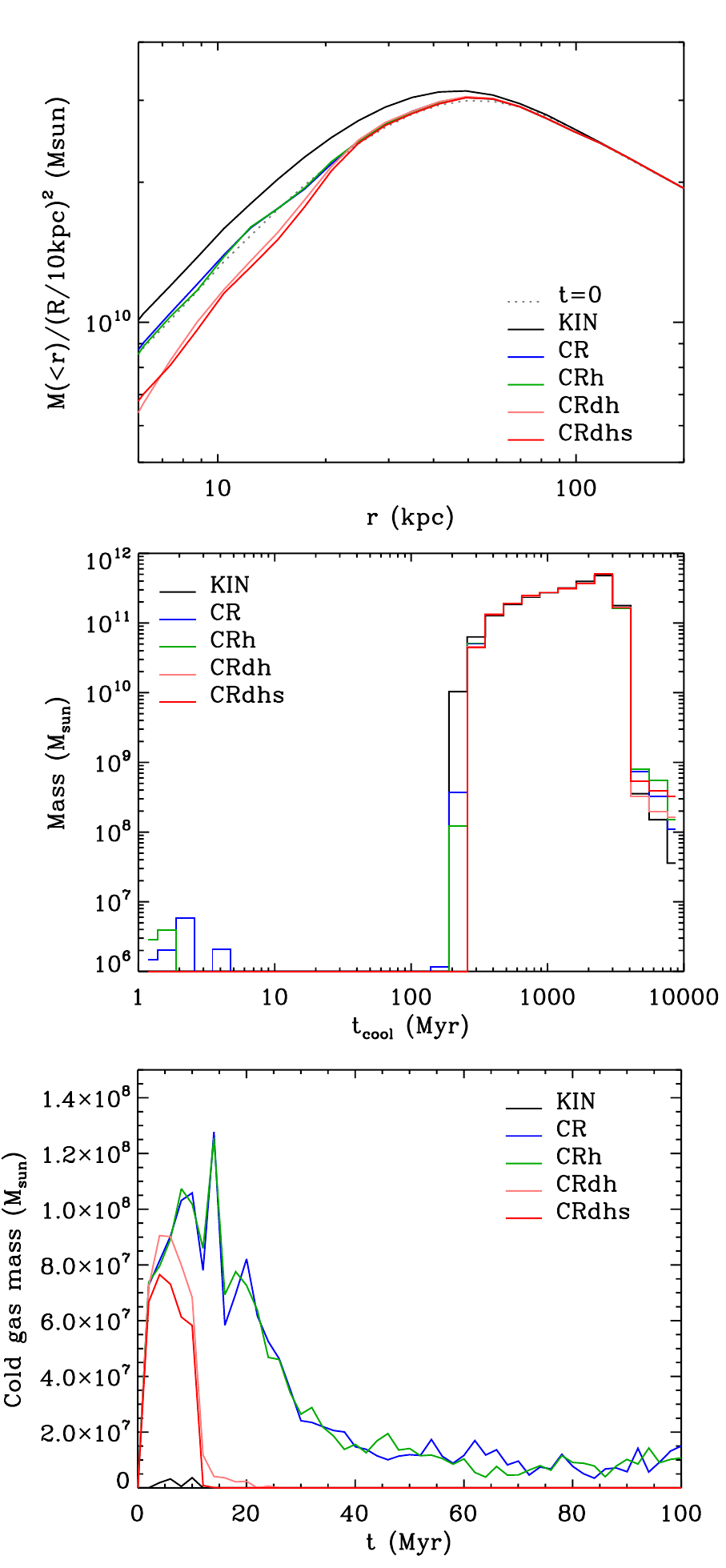} 
\caption{{\it Top:} Radial profiles of enclosed mass, normalized by $(r/10\ {\rm kpc})^2$ so that the variation can be more easily seen. The dotted line represents the initial enclosed mass profile. The profiles at $t=100$ Myr for the KIN, CR, CRh, CRdh, and CRdhs cases are shown in black, blue, green, pink, and red curves, respectively. {\it Middle:} Mass distributions of gas with different radiative cooling times at $t=100$ Myr. {\it Bottom:} Evolution of the cold gas mass for all cases. Note that data plotted in the figure is analyzed from the simulations including radiative cooling, which are counterparts to the default simulations listed in Table \ref{tbl:sims}.}
\label{fig:gas}
\end{center}
\end{figure}

The expansion of the hot ICM can be easily seen by looking at the profiles of enclosed mass (top panel in Figure \ref{fig:gas}). Figure \ref{fig:gas} shows the evolution for the hot and cold phases of the ICM. Since the amount of cold gas formed from local thermal instabilities is sensitively dependent on contention between heating and cooling, data plotted in this figure is analyzed from the set of simulations including radiative cooling, which are counterparts to the default simulations in Table \ref{tbl:sims}. For the set of simulations in which radiative cooling is not included (not shown here), the difference between the final and initial profiles indicate the ability for the AGN bubbles to lift up the ambient hot ICM. What we found is that for the kinetic jets, the enclosed mass profile at $t=100$ Myr is very close to the initial one, meaning that the uplift is very inefficient. For the simulations with CRs but no transport (CR and CRh runs), a larger amount of the hot ICM is pushed from the central $\sim 30$ kpc to larger distances compared to the KIN case. This is due to CR buoyancy as well as the large cross section of fat CR bubbles. For the simulations including additional CR transport (CRdh and CRdhs runs), the uplift is even more efficient because of additional expansion driven by CR heating. This same trend could also be seen for the radiative simulations shown in the top panel of Figure \ref{fig:gas}. But due to gas inflows caused by radiative cooling, the profiles at $t=100$ Myr for some of the simulations have greater values than the initial profile in the central tens of kpc.  

% Figs (profiles of Mtot, cell mass vs. tcool) to show although mass outflow from CR buoyancy reduces hot accretion, inefficient heating by CRs induces low entropy gas, leading to condensation 

The middle panel of Figure \ref{fig:gas} shows the mass distribution of gas with different cooling times at $t=100$ Myr. The distribution is bimodal: the left/right bump is from the cold/hot phase. Because of the more efficient uplift of the hot ICM by CRs, the amount of hot gas with short cooling times ($t_{\rm cool} \lesssim 250$ Myr) is significantly reduced in the simulations with CRs. Therefore, aside from the heating provided by the CRs, the more efficient driving of a hot-gas outflow can readily suppress the cooling flow. This dynamical impact on the hot phase by CRs has been also reported by \cite{Mathews08}. 

However, what has not been noted before is the formation of cold gas due to local thermal instabilities, which does not exist in the previous simulations possibly because they do not account for the self-consistent generation of bubbles by jets (and because of the absence of radiative cooling or the use of dropout terms). As can be seen from the middle and bottom panels of Figure \ref{fig:gas} (and also Figure \ref{fig:slices}), the amount of cold gas and its evolution are very different between the kinetic and CR jets and among simulations with different CR transport mechanisms. For the KIN simulation, very little cold gas is formed throughout the simulation because there is very efficient heating provided by the large amount of thermal energy contained with the bubbles via direct mixing \citep{Hillel16, Yang16b}. In contrast, for the runs with CRs but no CR transport (CR and CRh), a large amount of cold gas is formed in the first 20 Myr and there is still $\sim 10^7 M_\odot$ of cold gas at $t=100$ Myr. This is due to a combination of two effects. First, there is less heating because CR-jet-inflated bubbles contain less thermal energy for direct mixing (see Figure \ref{fig:prfs}). Second, there is stronger adiabatic cooling associated with the expansion of the ICM mentioned above.   

For the runs with CR transport (CRdh and CRdhs), there is also an initial episode of cold gas formation for similar reasons to the runs without CR transport (CR and CRh). However, the amount of cold gas is significantly reduced at $t=10-20$ Myr. That is because including CR transport due to diffusion (CRdh) or streaming (CRdhs) allows the CRs to escape from the bubbles and interact with the ambient ICM. Since Coulomb and hadronic heating are both proportional to the multiplication between CR number density and gas density, the heating is most efficient when the CRs could get into contact with the ICM. The additional heating due to streaming (the CRdhs run) results in less cold gas during the bubble formation and faster reduction of cold gas at $t=10-15$ Myr, but the overall evolution is similar to the CRdh run. Note that even though both the CRh and CRdh runs include collisional heating of CRs, they predict qualitatively different evolution of cold gas. This stresses the importance of CR transport in the process of ICM heating. 

In summary, we find that the evolution of the multiphase gas is very different for kinetic vs.\ CR jets and for simulations with and without CR transport mechanisms. In particular, CRs jets can drive more significant expansion of the hot ICM due to buoyancy and larger cross sections of the CR bubbles. Adiabatic cooling from this expansion, together with a lesser amount of heat provided by direct mixing, triggers an episode of cold gas formation during the bubble formation. The time evolution of the amount of cold gas is different for runs with and without CR transport processes -- cold gas is depleted because of heating enabled by the interaction between the CRs and the ambient ICM as a result of CR transport. 

Our result supports that of the simulations of self-regulated CR-jet feedback by \cite{R17}, who find that CR transport is crucial for establishing self-regulation when models include CRs\footnote{In purely hydrodynamic simulations (e.g., \citealt{Gaspari12, Yang16b}), a realistic duty cycle and self-regulation can be achieved without additional transport mechanisms.}. Without transport, the central SMBH in their simulation is active almost all the time and eventually inject too many CRs that violate the upper limit inferred by gamma-ray observations. This can be naturally explained by our result because without transport, the heating from CRs is very inefficient and the ICM is more prone to thermal instabilities due to more efficient uplift. Therefore, in this case the SMBH is continuously fed by cold gas without providing much heating to its surroundings. Another interesting result from their study is that the AGN activity is more episodic when CR jets are considered, compared to previous self-regulated simulations of kinetic jets \citep{Yang16b}. This is also consistent with our finding because CR jets are more capable of uplifting the hot ICM, and hence it would take a longer time for the gas to cool again and trigger another cycle of AGN activity.   

\subsection{Generation of Turbulence}
\label{sec:turb}

Since our simulations are carried out in 3D, it is meaningful to quantify the kinematics and compare the generation of turbulence by kinetic jets vs.\ CR jets. 

\begin{figure*}[htbp]
\begin{center}
\includegraphics[scale=0.75]{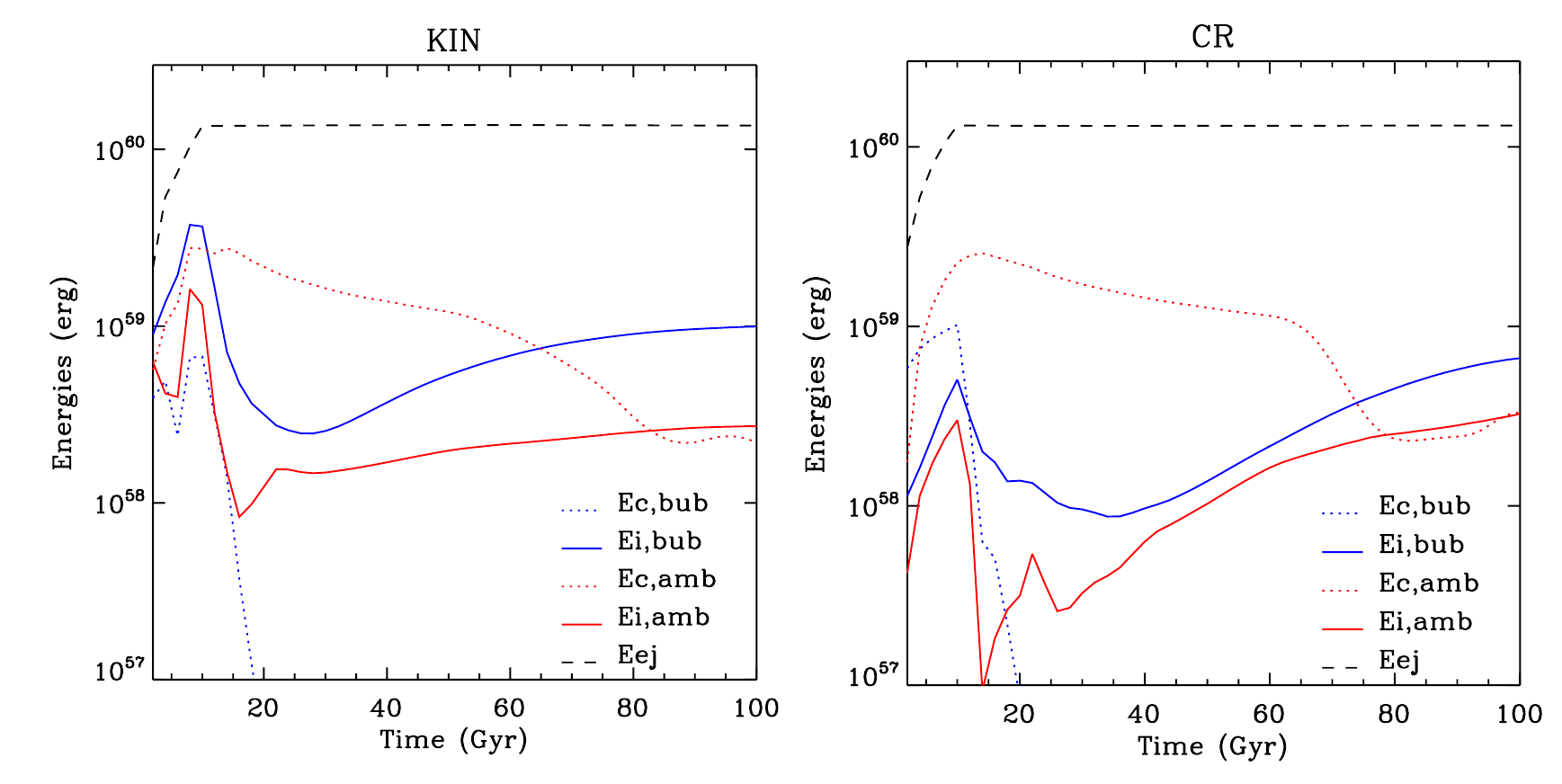} 
\caption{Evolution of the compressible (dotted line) and incompressible (solid line) components of the kinetic energy within $r=100$ kpc for the KIN (left) and CR (right) simulations. Blue and red colors represent values for the bubbles and the ambient ICM, respectively. For all cases, the kinetic energy contained in the incompressible mode, a proxy for the turbulent energy, is at the percent level compared to the total injected energy by the AGN (dashed line).}
\label{fig:ekin}
\end{center}
\end{figure*}

Figure \ref{fig:ekin} shows the evolution of the compressible (which traces shocks and sound waves) and incompressible (which measures turbulence and $g$-modes) components of the kinetic energy within the central 100 kpc for the KIN and CR cases. For later discussions we use the incompressible component as a proxy for the turbulent energy, though precisely it represents an upper limit. The decomposition of the velocity field is done following the method of \cite{Reynolds15} and \cite{Yang16a}. The other CR simulations are not shown here because they show similar evolution to the CR run. Blue and red curves represent values computed for the bubbles and the ambient ICM, respectively. As mentioned in \S\ref{sec:method}, a small fraction of CRs ($f_{\rm cr}=0.001$) is injected with the kinetic jets in order to dye the jet materials. Motivated by the initial ratio of $f_{\rm cr}$ between the kinetic jets and the CR jets, we define the bubbles to be regions where the CR energy density is larger then $10^{-13}$ and $10^{-10}$ erg cm$^{-3}$ for kinetic and CR jets, respectively. This choice is arbitrary, but our conclusion does not sensitively depend on the above thresholds. The total energy injected by the AGN jets are shown in the dashed lines.  

% Fig: Energies of compressible and incompressible components vs. t

The evolution of the compressible component in both cases is similar. The compressible mode within the bubbles is negligible except during the epoch of bubble formation. For the ambient gas, it is $\sim 20\%$ of the total injected energy by the AGN and gradually decreases with time. The drops at $t \sim 70$ Myr in both cases are caused by shocks that propagate outside the radius of 100 kpc. The turbulent energy for both the bubbles and the ambient gas, though having an initial spike at the early stage in the KIN case, remains at the percent level compared to the total injected energy of the AGN for both kinetic and CR jets. Specifically, the turbulent energy contained within the bubbles after $t=20$ Myr for both the KIN and CR cases is $\sim 1-6\%$ of the total injected energy from the AGN; this ratio is even smaller for the ambient ICM ($\sim 0.2-2.5\%$). This indicates that the generation of turbulence by AGN jets is `inefficient', i.e., confined to the low-level subsonic regime, consistently with previous findings \citep{Reynolds15, Yang16b, Bambic18, Gaspari18}. Interestingly, our results further show that this conclusion is independent of the bubble composition.  
It is important to note that, even though turbulence is subsonic ($\sigma_{\rm LOS}\sim 100$ km s$^{-1}$; ${\rm Mach \ll 1}$) and thus turbulent dissipation is negligible ($t_{\rm diss}\sim {\rm Mach}^{-2}\,t_{\rm eddy}$, with $t_{\rm eddy}=L/\sigma_v$ the eddy turnover time and $L$ the bubble injection scale), the action of the turbulent mixing is still a key component of the feedback process ($t_{\rm mix}\sim t_{\rm eddy}$) which helps to deposit the injected energy and to further promote thermal instability (see also \citealt{Gaspari17}).

\begin{figure*}[tbp]
\begin{center}
\includegraphics[scale=0.8]{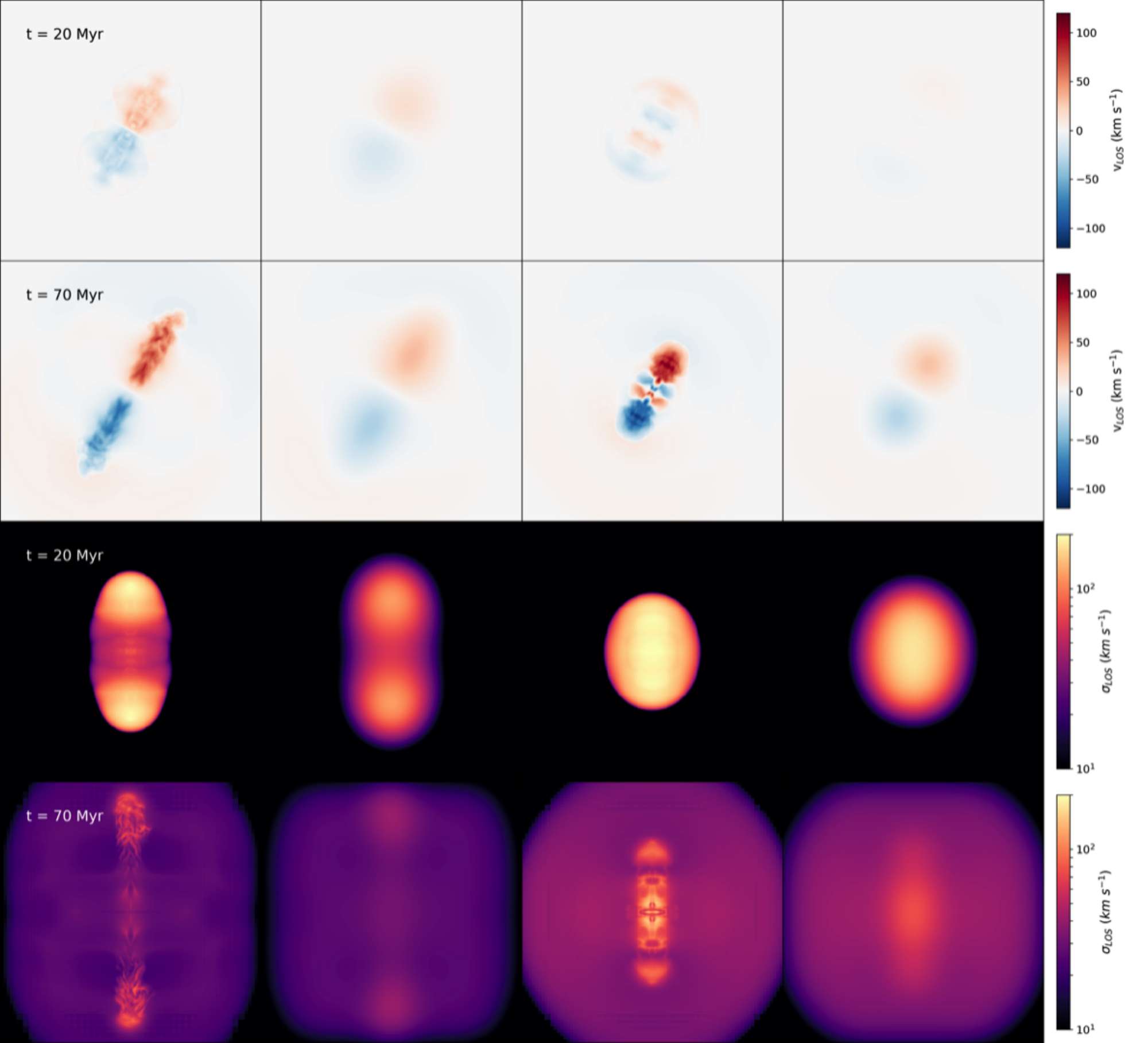} 
\caption{LOS velocities (upper two rows) and LOS velocity dispersions (bottom two rows) for the KIN (left two columns) and CR (right two columns) cases. For each simulation, two snapshots at $t=20$ Myr (first and third rows) and $t=70$ Myr (second and fourth rows) are shown with simulation resolution (first and third columns) and {\it Hitomi} resolution (second and fourth columns), respectively.}
\label{fig:hitomi}
\end{center}
\end{figure*}

In order to check consistency with the ICM kinematics in the Perseus cluster as observed by the {\it Hitomi} satellite \citep{Hitomi16, Hitomi18}, we compute maps of the line-of-sight (LOS) velocity, $v_{\rm LOS}=\left<v_l\right>$, and LOS velocity dispersion, $\sigma_{\rm LOS}=(\left<v_l^2 \right>-\left<v_l \right>^2)^{1/2}$, where $v_l$ is the velocity component along the LOS and brackets represent emission-weighted averages. Figure \ref{fig:hitomi} shows the maps for the KIN and CR simulations at $t=20$ and 70 Myr, with $v_{\rm LOS}$ and $\sigma_{\rm LOS}$ computed for both simulation resolution and {\it Hitomi} resolution of $\sim 20$ kpc (smoothed using a Gaussian filter to mimic the observed point spread function with a width of 20 kpc). The LOS velocity dispersion is projected along the $x$ axis in the simulation (perpendicular to the jet axis). For $v_{\rm LOS}$, we choose a LOS that is $45^\circ$ relative to the jet axis, because if one were to compute $v_{\rm LOS}$ along a direction either parallel or perpendicular to the jet axis, the value would be close to zero due to symmetry. Note that at $t=20$ Myr, the bubbles are just formed and the velocities plotted trace both the bubbles and the weak shocks; at $t=70$ Myr, the shocks already left the inner region. 

As can be seen from Figure \ref{fig:hitomi}, the values of $v_{\rm LOS}$ and $\sigma_{\rm LOS}$ are time-dependent for both KIN and CR simulations. At early stages of the bubble evolution, $v_{\rm LOS}$ is negligible ($\lesssim 20$ km s$^{-1}$) because the velocity field is highly symmetric. It then gradually increases with the time as the gas becomes more turbulent at later times (e.g., $v_{\rm LOS} \sim 50-100$ and $\sim 30-50$ km s$^{-1}$ at $t=70$ Myr for simulation resolution and {\it Hitomi} resolution, respectively). In contrast, $\sigma_{\rm LOS}$ is high ($\sim 150-200$ and $\sim 100-150$ km s$^{-1}$ for simulation and {\it Hitomi} resolutions, respectively) at early times due to both shocks and turbulence driven by the jets and decreases afterwards as the kinetic energy is spread out to a larger volume. It is non-trivial to make a direct comparison to the observed constraint obtained by {\it Hitomi} given uncertainties of the viewing angle, time since injection, and the fact that we only simulate a single AGN outburst. However, the range of the values is consistent with the observational constraint within the measurement uncertainties (at least at certain epoch and projection), indicating that AGN jets (aside from gas motions driven by other mechanisms such as cluster mergers \citep{Lau17,Roncarelli18} or substructures within the clusters \citep{Bourne17,Eckert17}) can be an important source for the detected bulk velocity and velocity dispersion by {\it Hitomi}. Finally, we note that turbulence in cluster cores is also built up in time during the recurrent hundreds cycles of CCA feeding/AGN feedback (cf., \citealt{Gaspari18}) that quench the cooling flow and preserve the CC for several billion years (at least since redshift $z\sim 2$; \citealt{McDonald17}).

% Fig: Maps of 1D velocity dispersions to compare with Hitomi data

\subsection{Observable signatures}

\subsubsection{Gamma-ray}
\label{sec:gamma}

An important observational constraint for any successful AGN feedback models based on heating from CRs is the non-detections of gamma-ray signals of observed clusters \citep[e.g.,][]{Fermi14, Fermi16}, which put stringent limits on the amount of CRs allowed within clusters. Although our simulations do not account for pre-existing CRs generated by structure formation shocks \citep[e.g.,][]{Pinzke10} or by subsequent AGN outbursts \citep[e.g.,][]{R17}, it is instructive to compute the expected gamma-ray fluxes for our simulated CRs, so as to make sure the simulated scenario does not demonstrably violate the observational constraints under realistic assumptions. 

For the computed gamma-ray fluxes, we have assumed a distance to the cluster same as the CC cluster A1795. For the following discussion, we will compare our results to the observed gamma-ray fluxes by {\it Fermi} above 500 MeV for A1795, which is $1.30\times 10^{-10}$ ph cm$^{-2}$ s$^{-1}$ \citep{Fermi14}. Note that this limit is obtained by assuming that the gamma rays originate from a point source. This is a good approximation since our simulated CRs are distributed only within the cluster core, which is much smaller than the diameter of the cluster ($\sim 4.04$ Mpc or an angular diameter of $\sim 0.95^\circ$) and the resolution of the {\it Fermi} satellite at 1 GeV ($\sim 1^\circ$). Note also that ideally one would like to compare to the observational limit for the Perseus cluster, which is what our initial condition is based upon. However, {\it Fermi} has found a significant contribution of the gamma-ray signal from the central galaxy NGC1275, which prohibits a limit to be placed for the diffuse ICM. For this reason we choose A1795, a nearby CC cluster with similar masses to the Perseus cluster.

We compute the predicted gamma-ray fluxes from our simulated CR bubbles in either the leptonic scenario, where gamma rays are produced by inverse-Compton (IC) scattering of photons in the cosmic microwave background (CMB) by CR electrons, or the hadronic scenario, in which the gamma rays are generated by inelastic collisions between CR protons and the thermal nuclei via the pion decay process. The detailed method is described in \cite{Yang13}. For the leptonic/hadronic scenario, we assume the simulated CR energy density is composed of purely CR electrons/protons, i.e., $f_{\rm e}\equiv e_{\rm cr,e}/e_{\rm cr}=1$ and $f_{\rm p}\equiv e_{\rm cr,p}/e_{\rm cr}=1$, respectively. For the hadronic model, the CR spectrum is assumed to have a spectral index of $-2$ and range from 1 GeV to 1 TeV, which could produce gamma rays between $\sim 100$ MeV and $\sim 100$ GeV. For the leptonic model, we assume that the CR spectrum ranges from 1 GeV to 1 TeV and has a spectral index of $-2.1$. The lower-energy threshold is somewhat arbitrary (as long as it is smaller than the energy of CRe that is required to IC scatter the CMB photons to the observed energy of 500 MeV). The higher-energy cutoff of the CR electron spectrum is expected to be strongly time-dependent due to IC and synchrotron cooling. The choice of 1 TeV, though reasonable for younger jets, is likely an over-estimate for older bubbles. However, the computed gamma-ray fluxes primarily come from lower-energy CRs and hence do not sensitively depend on the assumption of the high-energy cutoff. We note that the above assumptions about the CR spectrum would not be needed for simulations that evolve the CR spectrum self-consistently \citep[e.g.,][]{Yang17}; we will explore this in future work. Since the resultant gamma-ray flux for each simulation is largest after the initial injection and monotonically decreases with time, we quote two epochs, $t=10$ and 100 Myr, in order to bracket the variation due to the bubble evolution. For all calculations, we assume the area extended by the CRs as viewed from the $x$ axis (perpendicular to the jets). 

Our results are summarized in Table \ref{tbl:gamma}. Assuming the hadronic scenario, all gamma-ray fluxes are at least a factor of $\sim 40$ below the observed limit. The low level of the predicted hadronic emission is not surprising because of the long characteristic timescales for hadronic interactions ($>$ Gyr for typical cluster parameters). Since our simulations only include a single AGN outburst, this leaves plenty of room for CRs produced by multiple injections. Therefore, even if all of the simulated CRs are protons, our models do not have apparent contradictions with the current {\it Fermi} constraints. 

In the leptonic scenario, on the other hand, while the low CR fraction within the kinetic jets allow the KIN simulations to have gamma-ray fluxes below the observed limits, the CR jets produce gamma rays that are $\sim 30-50$ times more than the {\it Fermi} constraint. This means that the simulated CRs cannot be all composed of CR electrons, i.e., $f_{\rm e}\ll 1$. In fact, by requiring the predicted fluxes to be below the observed value, we can derive an upper limit for the CR electron fraction within the jets to be $f_{\rm e} < 0.022$. This constraint would become even more stringent if we were to account for CR electrons from multiple episodes of injections that have not aged. Although the estimate done here is rather crude, interestingly the conclusion that CR electron population does not dominate the bubble pressure is consistent with compositions inferred from many observed radio galaxies and cluster radio bubbles \citep[e.g.,][]{Dunn04, DeYoung06, Birzan08, Croston18}. The estimate also implies that the dynamics in the system is dominated by CRp rather than CRe, justifying our treatment of CRs as purely CRp in the simulations.   

% Fig: Gamma-ray fluxes assuming hadronic and leptonic models

\capstartfalse
\begin{deluxetable*}{ccccccc}
\tabletypesize{\normalsize}
\tablewidth{0.9\textwidth}
\tablecaption{Predicted gamma-ray fluxes above 500 MeV}
%\hline\hline
\tablehead{
\vspace{0.3cm} \\
\colhead{Mechanism} & 
\colhead{Run} & 
\colhead{$t$} & 
\colhead{Intensity} & 
\colhead{Area} & 
\colhead{Solid angle} & 
\colhead{Flux} \\
\colhead{ } & 
\colhead{ } & 
\colhead{(Myr)} & 
\colhead{(ph s$^{-1}$ cm$^{-2}$ sr$^{-1}$)} & 
\colhead{(kpc$^2$)} & 
\colhead{(sr)$^{({\rm a})}$} & 
\colhead{(ph s$^{-1}$ cm$^{-2}$)$^{({\rm b})}$}
}
\startdata
\vspace{+0.25cm} \\
Hadronic & KIN & 10 & $6.53\times 10^{-8}$ & 1202 & $1.97\times 10^{-8}$ & $1.29\times 10^{-15}$\\
Hadronic & KIN & 100 & $6.55\times 10^{-8}$ & 4916 & $8.06\times 10^{-8}$ & $5.28\times 10^{-15}$\\
Hadronic & CR & 10 & $3.32\times 10^{-5}$ & 1670 & $2.74\times 10^{-8}$ & $9.10\times 10^{-13}$\\
Hadronic & CR & 100 & $3.63\times 10^{-5}$ & 5754 & $9.43\times 10^{-8}$ & $3.42\times 10^{-12}$\\
Leptonic & KIN & 10 & $3.85\times 10^{-4}$ & 1202 & $1.97\times 10^{-8}$ & $7.58\times 10^{-12}$\\
Leptonic & KIN & 100 & $5.80\times 10^{-5}$ & 4916 & $8.06\times 10^{-8}$ & $4.67\times 10^{-12}$\\
Leptonic & CR & 10 & $2.19\times 10^{-1}$ & 1670 & $2.74\times 10^{-8}$ & $6.00\times 10^{-9}$\\
Leptonic & CR & 100 & $3.80\times 10^{-2}$ & 5754 & $9.43\times 10^{-8}$ & $3.58\times 10^{-9}$
\enddata
\vspace{+0.3cm}
\tablenotetext{}{\small {\it Notes.} 
(a) The solid angle is calculated as $\Omega=A/D^2$, where $A$ is the area extended by the CRs and $D$ is the angular diameter distance to the cluster. The values quoted in the table is computed assuming distance to the A1795 cluster, for which $D=247.0$ Mpc. 
(b) The gamma-ray fluxes are computed in the leptonic/hadronic scenario assuming the simulated CR energy density is composed of purely CR electrons/protons. The predicted fluxes are to be compared to the observed {\it Fermi} upper limits for A1795, $1.30\times 10^{-10}$ ph s$^{-1}$ cm$^{-2}$ \citep{Fermi14}.  
\\}
\label{tbl:gamma}
\end{deluxetable*}
\capstarttrue

\subsubsection{Sunyaev-Zel'dovich effect}
\label{sec:sz}

The composition of cluster radio bubbles is still largely unknown. Since the total pressure of many observed bubbles is much higher than the pressure contributed by CR electrons \citep[e.g.,][]{Dunn04}, the bubbles must be dominated by magnetic pressure, ultra-hot thermal plasma, or CR protons. One of the proposed methods to distinguish AGN bubbles dominated by ultra-hot thermal gas and CRs is the thermal Sunyaev-Zel'dovich (SZ) effect \citep{SZ72, Birkinshaw99}, which arises because the CMB photons are IC-scattered by electrons of the ICM and are spectrally redistributed. Since it is one of the most promising observational techniques for distinguishing thermal-energy- and CR-dominated bubbles, in this section we generate synthetic images of the SZ effect for the three representative runs, i.e., KIN, CR, and CRdh simulations. 

\begin{figure}[tbp]
\begin{center}
\includegraphics[scale=0.43]{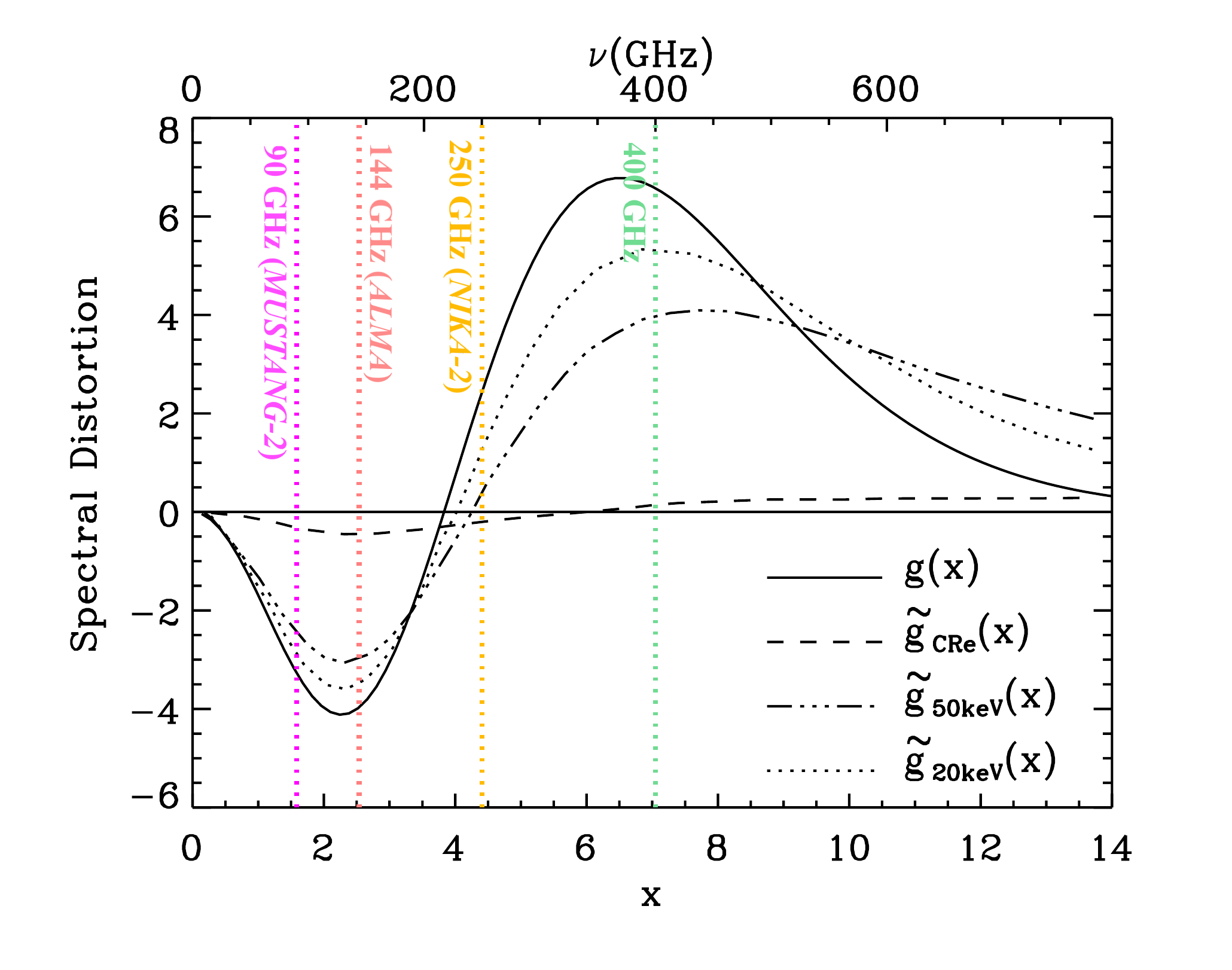} 
\caption{Spectral distortions as a function of the dimensionless frequency, $x=h\nu/k_{\rm B}T_{\rm CMB}$, due to the thermal SZ effect $g(x)$, relativistic SZ effect due to a population of power-law CR electrons, $\tilde{g}_{CRe}=[j(x)-i(x)]\tilde{\beta}_{CRe}$, and the relativistic SZ effect due to a population of ultra-hot thermal electrons, $\tilde{g}_{50keV}=[j(x)-i(x)]\tilde{\beta}_{th}(50$ keV), and due to thermal electrons with $kT_{\rm e}=20$ keV, respectively. The vertical lines from left to right show frequencies of 90 GHz, 144 GHz, 250 GHz, and 400 GHz.}
\label{fig:szgx}
\end{center}
\end{figure}

We follow the method of \cite{Pfrommer05} to compute the relative change in flux density as a function of dimensionless frequency, $x=h\nu/(kT_{\rm CMB})$ ($k$ is the Boltzmann constant), 
\begin{equation}
\delta i(x) = g(x) y_{\rm gas} [1+\delta(x,T_{\rm e})] + \delta i_{\rm rel}(x)
\end{equation}
The first term is the contribution from non-relativistic electrons, with the spectral distortion and the Compton $y$ parameter given by, respectively, 
\begin{eqnarray}
g(x) &=& \frac{x^4 e^x}{(e^x-1)^2} \left( x \frac{e^x+1}{e^x-1} -4 \right), \\
y_{\rm gas} &=& \frac{\sigma_T}{m_{\rm e}c^2} \int dl n_{\rm e,gas} kT_{\rm e},
\end{eqnarray}
where $\sigma_{\rm T}$ is the Thompson cross section, $m_{\rm e}$ is the electron rest mass, $c$ is the speed of light, $n_{\rm e,gas}$ and $T_{\rm e}$ are thermal electron number density and temperature, respectively, and $\delta(x,T_{\rm e})$ is the relativistic correction term \citep[e.g.,][]{Ensslin00}. The second term represents spectral distortions owing to relativistic electrons, 
\begin{eqnarray}
\delta i_{\rm rel}(x) &=& [j(x)-i(x)] \tau_{\rm rel} = \tilde{g} (x) \tilde{y},\\
\tilde{y} &=& \frac{\sigma_T}{m_{\rm e}c^2} \int dl n_{\rm e} k\tilde{T}_{\rm e},\\
k\tilde{T}_{\rm e} &=& \frac{P_{\rm e}}{n_{\rm e}}, \label{eq:kTe}\\ 
\tilde{g} (x) &=& [j(x)-i(x)] \tilde{\beta}(k\tilde{T}_{\rm e}),\\
\tilde{\beta}(k\tilde{T}_{\rm e}) &=& \frac{m_{\rm e}c^2 \int dl n_{\rm e}}{\int dl n_{\rm e} k \tilde{T}_{\rm e}}.
\end{eqnarray}
Here $\tau_{\rm rel}=\sigma_{\rm T} \int dl n_{\rm e}$ is the optical depth of Compton scattering with relativistic electrons, $i(x)=x^3/(e^x-1)$ describes the frequency distribution of the CMB, $i(x)\tau_{\rm rel}$ is the flux scattered to other frequencies, $j(x)\tau_{\rm rel}$ is the flux scattered from other frequencies to $x$, $\tilde{\beta}(k\tilde{T}_{\rm e})$ is the normalized pseudo-thermal beta parameter, and $k\tilde{T}_{\rm e}$ is the pseudo temperature of the relativistic electrons. The scattered flux can be expressed as
\begin{equation}
j(x) = \int_0^\infty dt \int_0^\infty dp f_{\rm e}(p) P(t;p) i(x/t),
\end{equation}
where $P(t;p)$ is the photon redistribution function for electrons of normalized momentum $p=\beta_{\rm e}\gamma_{\rm e}$ ($\beta_{\rm e}\equiv v/c$ and $\gamma_{\rm e} \equiv 1/\sqrt{1-\beta^2}$), $f_{\rm e}(p)$ is the electron spectrum normalized to 1, and the frequency of a scattered photon is shifted by a factor of $t$. The photon redistribution can be derived in the Thomson regime ($\gamma_{\rm e} h\nu \ll m_{\rm e}c^2$; e.g., \citealt{Ensslin00}):
\begin{eqnarray}
P(t;p) = &-& \frac{3|1-t|}{32p^6t} \left[ 1+(10+8p^2+4p^4)t+t^2 \right] \nonumber \\
&+& \frac{3(1+t)}{8p^5} \Bigg\{ \frac{3+3p^2+p^4}{\sqrt{1+p^2}} \nonumber \\
&-& \frac{3+2p^2}{2p} \left[ 2 {\rm arcsinh}(p) - |\ln(t)| \right]
\end{eqnarray}
for $|\ln(t)| \leq 2$arcsinh$(p)$ and $P(t;p)=0$ otherwise. 
Note that $j(x)$ depends on the electron spectrum $f_{\rm e}(p)$ and hence the spectral distortion $\tilde{g}(x)$ is different for ultra-hot thermal plasma and CRs. Figure \ref{fig:szgx} plots the SZ spectral distortions as a function of frequency for four illustrative cases: (1) non-relativistic thermal plasma; (2) a power-law distribution of CR electrons with
\begin{equation}
f_{\rm CRe}(p,\alpha,p_1,p_2)=\frac{(\alpha-1)p^{-\alpha}}{p_1^{1-\alpha}-p_2^{1-\alpha}},
\end{equation}
where $(p_1,p_2)=(1,10^3)$ and $\alpha=2$ are assumed for the spectral range and spectral index, respectively; (3) a population of ultra-hot thermal electrons with $kT_{\rm e}=50$ keV:
\begin{equation}
f_{\rm e,th}(p,\beta_{\rm th})=\frac{\beta_{\rm th}}{K_2(\beta_{\rm th})} p^2 \exp \left( -\beta_{\rm th} \sqrt{1+p^2} \right),
\end{equation}
where $\beta_{\rm th}=m_{\rm e}c^2/(kT_{\rm e})$ is the normalized thermal beta-parameter, and $K_2$ is the modified Bessel function of the second kind; (4) ultra-hot thermal electrons with $kT_{\rm e}=20$ keV. The SZ spectral distortions caused by ultra-hot thermal gas are somewhat closer to that of the thermal gas, whereas CR electrons result in a  very small spectral distortion. This is why the thermal SZ was proposed as a promising means to distinguish ultra-hot thermal bubbles from CR bubbles \citep{Pfrommer05}.

For our simulated cluster, we calculate $\delta i(x)$ assuming the LOS is along the $x$ axis. For each LOS, we sum over contributions from the thermal gas (defined as gas with $kT_{\rm e}<20$ keV), ultra-hot thermal plasma ($kT_{\rm e}\geq 20$ keV), and CRs. For the CR component, the pressure of relativistic electrons in Eq.\ \ref{eq:kTe} is directly related to the simulated CR energy density by $P_{\rm e}=f_{\rm e} P_{\rm cr} = f_{\rm e} (\gamma_{\rm cr}-1) e_{\rm cr}$, where $f_{\rm e}$ is the fraction of CR electron pressure to the total CR pressure (see \S\ref{sec:gamma}). If $f_{\rm e} \ll 1$, the contribution from CRs to the SZ signal would be negligible, maximizing the difference between CRs and thermal gas. While this is a likely scenario (as discussed in \S\ref{sec:gamma}) and optimal for distinguishing thermal and CR bubbles, to be conservative we assume $f_{\rm e}=0.5$ (equipartition between CR electrons and protons). Using a smaller $f_{\rm e}$ would not significantly affect the results because the spectral distortion caused by CRs is already very small (see Figure \ref{fig:szgx}). Indeed, we verified that using $f_{\rm e}=0.01$ results in negligible changes in the results presented below.

\begin{figure*}[tbp]
\begin{center}
\includegraphics[scale=0.9]{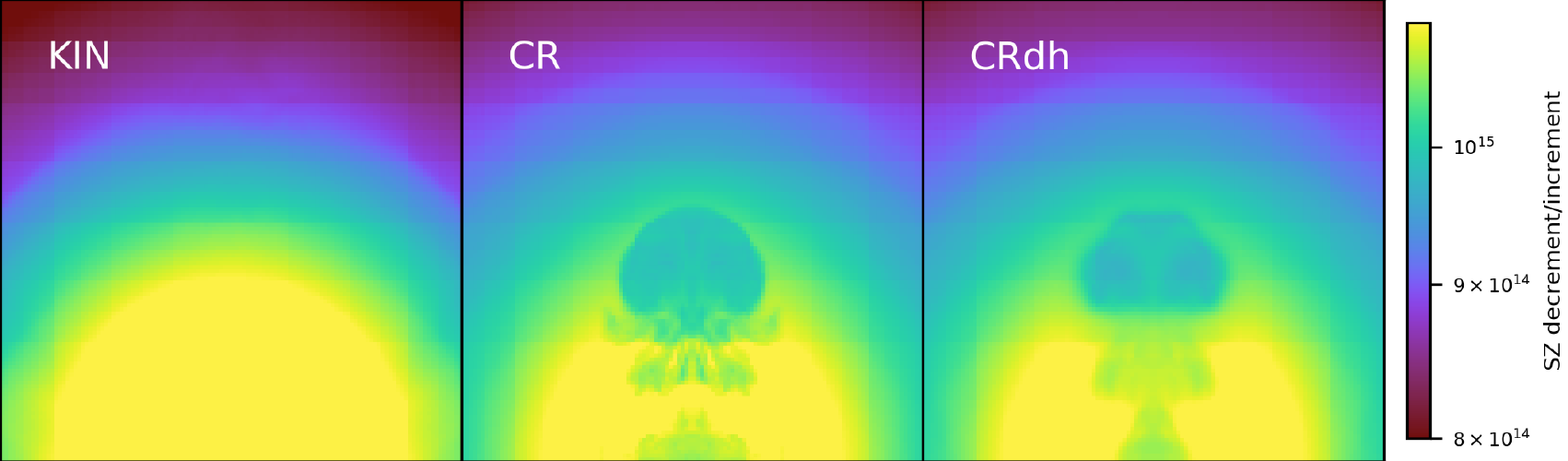} 
\caption{Absolute values of the simulated SZ decrement (in arbitrary units) at $t=50$ Myr for the KIN (left), CR (middle), and CRdh (right) simulations at 144 GHz (the frequency on which ALMA Band 4 is centered). It is apparent that the CR-dominated bubbles show deficits in the SZ signal similar to the X-ray cavities (see Figure \ref{fig:slices}), whereas ultra-hot thermal bubbles in the KIN case do not show suppression in the SZ image.}
\label{fig:sz}
\end{center}
\end{figure*}

Figure \ref{fig:sz} shows the SZ decrement for the KIN, CR, and CRdh simulations at $t=50$ Myr at 144 GHz (on which ALMA Band 4 is centered). From the synthetic SZ maps, one can immediately see the distinct visual appearances between bubbles dominated by ultra-hot thermal gas (as in the KIN case) and bubbles dominated by CRs (CR and CRdh cases). In the latter case, suppression of the SZ flux density can be clearly seen for sightlines passing through the CR bubbles, resembling the X-ray cavities (see Figure \ref{fig:slices}). The CR and CRdh cases look similar, though the SZ bubbles in the CRdh run have smoother edges due to CR diffusion. In contrast, the ultra-hot thermal bubbles are essentially indistinguishable from the thermal gas in their surroundings.

\begin{figure}[tbp]
\begin{center}
\includegraphics[scale=0.55]{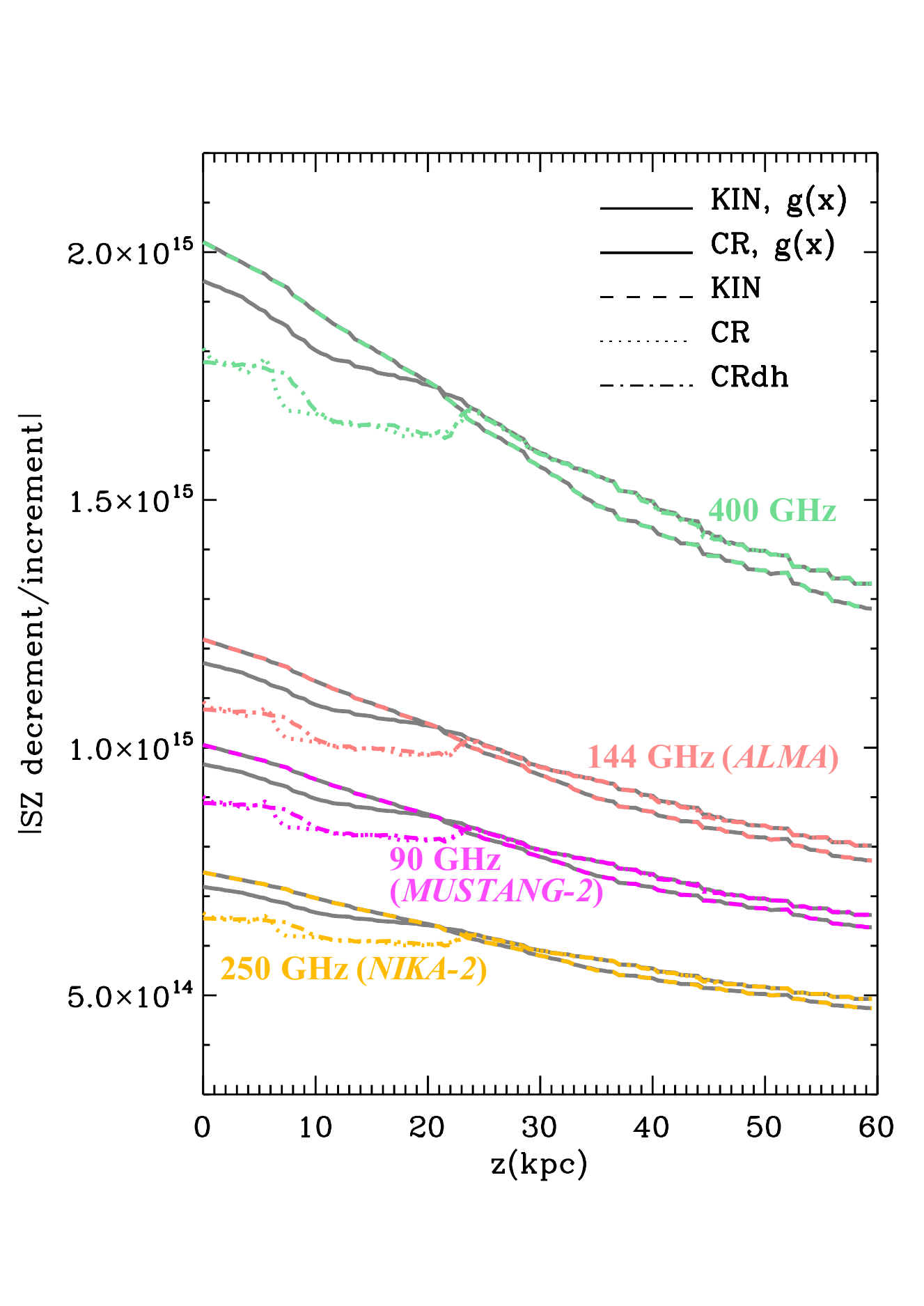} 
\caption{Profiles along the z-axis of the projected SZ decrement/increment (in absolute magnitudes) for the KIN, CR, and CRdh cases. Curves of different colors represent four characteristic frequencies, namely, 90 GHz, 144 GHz, 250 GHz, and 400 GHz. For both the thermal and CR bubbles, we overplot the control cases in which we assume the bubbles are filled with non-relativistic thermal gas (gray solid lines). The CR bubbles show clear deficits in the SZ signal, whereas the ultra-hot thermal bubbles in the KIN case show negligible suppression.}
\label{fig:szprf}
\end{center}
\end{figure}

To be more quantitative, in Figure \ref{fig:szprf} we plot vertical profiles of the projected SZ decrement/increment for the KIN, CR, and CRdh cases at four characteristic frequencies, namely, 90 GHz (relevant for MUSTANG-2), 144 GHz (one of ALMA bands), 250 GHz (one of the NIKA-2 frequency bands), and 400 GHz. For both the thermal and CR bubbles, we overplot the control cases in which we assume the bubbles are filled with non-relativistic thermal gas (gray solid lines). In these control cases, there is a smooth transition from sightlines passing through the bubbles to those through the ambient gas. The magnitude of the signals at different frequencies directly reflects the expected amount of spectral distortions (Figure \ref{fig:szgx}). For all frequencies, it is evident that the CR bubbles (both in the CR and CRdh cases) show suppression of the SZ signal for sightlines passing through the bubbles ($|z| \lesssim 23$ kpc). Overall, the amount of suppression is $\sim$\,$6-9\%$ with respect to the control case, consistent with previous estimates \citep{Pfrommer05}.

In contrast, the ultra-hot thermal bubbles show essentially no deficits in the SZ signals compared to the control case. This is owing to a few effects: (1) the SZ spectral distortion of ultra-hot thermal gas, though having a smaller amplitude, is similar to that of non-relativistic thermal gas (see Figure \ref{fig:szgx}); (2) any contrast between the bubbles and the ambient medium is diluted by LOS projections because of the small path length of the bubbles ($\sim 10-20$ kpc); (3) the bubbles in the KIN case at $t=50$ Myr, which are self-consistently generated by kinetic-energy-dominated jets of typical parameters, have a maximum and average temperature of $\sim 25.6$ keV and $\sim 8.2$ keV, respectively. This could explain why the calculation by \cite{Pfrommer05}, which assumes a uniform temperature distribution for the ultra-hot thermal bubbles at 20 keV and 50 keV, shows a small but nonzero suppression, while our results show negligible deficits by considering a more realistic temperature distribution within the bubbles. 

% Fig: SZ maps

Our results confirm previous suggestion that SZ observations of cluster radio bubbles with high-resolution, high-sensitivity SZ instruments are key to setting constraints on the bubble composition. In fact, a recent study by \cite{Abdulla18} has applied this method to the large cavities in MS 0735.6+7421. They claim detections of the SZ deficits coincident with the X-ray cavities (although with low S/N), which allows them to conclude that the cavities are supported either by thermal plasma with temperature greater than hundreds of keV or by CRs. Future SZ observations of other cavities, in particular coming from MUSTANG-2 and NIKA-2 ground telescopes will be instrumental in unveiling the content of AGN bubbles with high significance.

%===================================================

\section{Conclusions}
\label{sec:conclusion}

Investigating feeding and feedback of the central SMBH is key to understanding the dynamics and thermodynamics of the ICM in the cores of galaxy clusters. Bubbles inflated by AGN jets could stir up the gas, provide heat to the ICM to counteract radiative cooling globally, and could trigger cold-gas condensation due to local thermal instabilities. While kinetic-energy-dominated jets have been extensively studied using purely hydrodynamic simulations, the effects of CR-dominated jets are less well understood. To this end, we perform 3D hydrodynamic simulations of CR-dominated jets in a Perseus-like cluster to study the detailed evolution of a single AGN outburst. In particular, we focus on their impact on the process of heating and cooling, the generation of turbulence, and the observable signatures. We contrast CR-dominated jets with kinetic-energy-dominated jets, and we compare simulations with and without CR transport processes. Our main results are as follows. 

1.\ By injecting jets with different energy partitions in kinetic and CR forms while keeping jet momentum the same, we confirm that kinetic-jet inflated bubbles tend to be more elongated, whereas fatter bubbles such as the young cavities observed at the center of the Perseus cluster are more easily produced by CR-dominated jets.  

2.\ CR bubbles can drive a more significant expansion of the hot ICM due to buoyancy and larger cross sections, which helps to suppress radiative cooling by removing gas with short cooling times near the cluster center. Since it takes longer times for the ICM to cool again and feed the SMBH, this effect could explain the more episodic AGN activity seen in previous simulations of self-regulated CR-jet feedback.

3.\ Heating by CR jets is less efficient than kinetic jets because less thermal energy is contained within the CR bubbles that could be accessed by the ICM through direct/turbulent mixing. The inefficient heating, together with adiabatic cooling associated with the expansion of the atmosphere, induces episodes of cold-gas formation during the bubble formation. This condensed multiphase gas is later crucial for the triggering of the AGN via CCA, which is the main agent of the feedback self-regulation.

4.\ The evolution of the cold gas sensitively depends on whether CR transport mechanisms are included or not. With transport by either diffusion or streaming, the CRs could escape the bubbles and interact with the ICM, thereby providing heating and greatly reducing the amount of cold gas at later times. This could explain why, in previous self-regulated CR-feedback simulations, the cluster can reach self-regulation only when CR transport processes are included. 

5.\ We show that the generation of turbulence by AGN jets (at least for one event) is mild, regardless of the jet composition. Indeed, for both kinetic and CR jets, the turbulent energy is at the percent level (no more than $\sim 6\%$) compared with the total injected energy of the AGN jets, which disfavors turbulent dissipation (though not mixing) as the primary heating mechanism.
Nevertheless, the low level of velocity dispersion and bulk motions are consistent within uncertainties with the {\it Hitomi} measurements (of order of 100 km s$^{-1}$). Multiple generation of bubbles, together with cosmic flows, may however augment such turbulence, as shown in other simulations.

6.\ We calculate the predicted gamma-ray emission from the simulated CR bubbles. Assuming the hadronic scenario, all estimates are below the current observed limits provided by {\it Fermi}. Assuming the leptonic model, the predicted gamma-ray fluxes from the CR-dominated jets are $\sim 30-50$ times above the observed limit, which allows us to put constraints on the fraction of CR electrons within the jets to be below $\sim 2.2\%$.

7.\ We generate synthetic SZ images and profiles at multiple frequencies and find that bubbles dominated by ultra-hot thermal plasma (as inflated by kinetic jets) present a negligible contrast relative to their surroundings, whereas CR-dominated bubbles show a clear deficit ($\sim 6-9\%$) in the SZ signal, similar to the X-ray cavities. We confirm previous suggestion that high-resolution, high-sensitivity SZ observations are a powerful tool for constraining the composition of cluster radio bubbles. Overall, this work will be key for the current/next-generation SZ observations performed with revolutionary SZ telescopes as ALMA, MUSTANG-2, and NIKA-2.

%===================================================

\section*{\bf \scriptsize Acknowledgements}
The authors thank Ming Sun for contribution to the early development of this project. HYKY acknowledges support from NASA ATP (grant number NNX17AK70G) and NSF grant AST 1713722.
MG is supported by NASA through Einstein Postdoctoral Fellowship Award Number PF5-160137 issued by the Chandra X-ray Observatory Center, which is operated by the SAO for and on behalf of NASA under contract NAS8-03060. Support for this work was also provided by Chandra GO7-18121X. 
The simulations presented in this paper were performed on {\tt Pleiades}, provided by the NASA High-End Computing (HEC) Program through the NASA Advanced Supercomputing (NAS) Division at Ames Research Center, as well as the {\tt Deepthought2} cluster supported by the Division of Information Technology at the University of Maryland College Park. FLASH was developed largely by the DOE-supported ASC/Alliances Center for Astrophysical Thermonuclear Flashes at University of Chicago. Data analysis presented in this paper was conducted with the publicly available $yt$ visualization software \citep{yt}. We are grateful to the $yt$ development team and community for their support. 
We acknowledge the `Multiphase AGN Feeding \& Feedback'\footnote{\href{http://www.sexten-cfa.eu/event/multiphase-agn-feeding-feedback/}{www.sexten-cfa.eu/event/multiphase-agn-feeding-feedback}} workshop at Sexten CfA (Italy) for stimulating interactions that helped to improve this work.
%===================================================

\bibliographystyle{biblio}
\bibliography{agn}

\end{document}